\documentstyle[12pt]{report}

\begin{document}

\pagestyle{empty}

\title{\LARGE{\bf{Fermionic counting of $\sf RSOS$-states and \\
Virasoro character formulas for \\
the unitary minimal series $\sf M(\nu ,\nu +1)$.\\
Exact results. }}}

\author{\bf{Alexander Berkovich}\thanks{berkovic@pib1.physik.uni-bonn.de} \\
{\em Physikalisches Institut der} \\
{\em Rheinischen Friedrich-Wilhelms Universit\"{a}t Bonn} \\
{\em Nussallee 12} \\
{\em D-53115 Bonn, Germany}}

\date{March 14, 1994 \\
Revised August 24, 1994 \\
\footnotesize{\em Dedicated to the memory of Mikhail Nirenberg}}

\maketitle

\begin{abstract}

\noindent
The Hilbert space of an $RSOS$-model, introduced by Andrews, Baxter, and
Forrester, can be viewed as a space of sequences (paths) $\{a_0,a_1,\ldots,
a_L\}$, with $a_j$-integers restricted by
$1\leq a_j\leq \nu,\;\;\mid a_j-a_{j+1}\mid=1,\;\;a_0\equiv s,\;\;
a_L\equiv r$.
In this paper we introduce different basis which, as shown here, has the same
dimension as that of an $RSOS$-model. This basis appears naturally in the
Bethe ansatz calculations of the spin $\frac{\nu-1}{2}$ $XXZ$-model.
Following McCoy et al, we call this basis -- fermionic (FB). \\
Our first theorem $Dim(FB)=Dim(RSOS-basis)$ can be succinctly expressed in
terms of some identities for binomial coefficients. Remarkably, these
binomial
identities can be $q$-deformed. Here, we give a simple proof of these
$q$-binomial identities in the spirit of Schur's proof of the Rogers-Ramanujan
identities. Notably, the proof involves only the elementary recurrences for
the $q$-binomial coefficients and a few creative observations. \\
Finally, taking the limit $L\rightarrow \infty$ in these $q$-identities, we
derive an expression for the character formulas of the unitary minimal
series $M(\nu,\nu+1)$ {\bf "Bosonic Sum $\equiv$ Fermionic Sum"}.
Here, Bosonic Sum denotes Rocha-Caridi representation
($\chi_{r,s=1}^{\nu,\nu+1}(q)$)
and Fermionic Sum stands for the companion representation recently conjectured
by the McCoy group \cite{3}.

\end{abstract}

\pagestyle{plain}

\section*{1. Introduction}

\setcounter{chapter}{1}
\setcounter{section}{0}
\setcounter{equation}{0}

The last decade has witnessed a remarkable convergence of ideas in such
diverse areas of mathematical physics as theory of knots and links,
classical and quantum exactly integrable systems, two-dimensional
gravity, string and conformal field theories and others. The emerging
mathematical structure, however, makes one turn around and examine
number theoretical questions which often date back to the time of
ancient Greece. \\
In a very interesting development \cite{1}-\cite{6}, celebrated identities
of the Rogers-Ramanujan type were revisited and many new $q$-series
for the characters of irreducible highest weight representations of
conformal field theories (CFT) were proposed. As a result, numerous
new links between Bethe ansatz approach \cite{7}, CFT, modular
forms, Rogers dilogarithmic functions and geometry of flag manifold
were established \cite{1}-\cite{6}, \cite{8}-\cite{12}. \\
In particular, Rogers-Ramanujan identities \cite{13}
\begin{eqnarray}
\sum_{m=0}^{\infty}\frac{q^{m(m+r)}}{(q,q)_m} & = & \frac{1}{(q,q)_{\infty}}
\cdot\sum_{j=-\infty}^{\infty}\{q^{j(10j+1+2r)}-q^{(2j+1)(5j+2-r)}\}\;;
\;\;\;\;\; r=0,1 \nonumber \\
(q,q)_m & \equiv & \prod_{i=1}^m(1-q^i)
\label{i1}
\end{eqnarray}

\noindent
can be expressed in the modern language as "Fermionic Sum $\equiv$
Bosonic Sum" representation of characters of the non-unitary minimal
model $M(2,5)(c=-\frac{22}{5})$, with $r=1(0)$ being a label for
a primary field of conformal dimension $0(-\frac{1}{5})$.
We also mention that the character itself can be thought as a
"$q$-dimension" of the Bethe ansatz basis. \\
In this paper we consider the following generalization of (\ref{i1})
related to $M(\nu,\nu+1)$ conformal series $(c=1-\frac{6}{\nu(\nu+1)})$
\[
\sum^{\infty}_{\stackrel{m_1,m_2,\ldots,m_{\nu-2}=0}
{m_i\equiv V_{i,r}^{\pm}(mod 2)}}
\frac{q^{\frac{\vec{m}^tC\vec{m}}{4}}}{(q,q)_{m_{\nu-2}}}
\prod_{i=1}^{\nu-3} \left[ \begin{array}{c}
\frac{1}{2}(K_{\nu-2}\cdot\vec{m})_i+\frac{1}{2}\delta_{i,b^{\pm}} \\ m_i
\end{array}\right]_q =
\]
\begin{equation}
\frac{q^{\frac{r(r-1)}{4}}}{(q,q)_{\infty}}\sum_{j=-\infty}^{\infty}
(q^{j[j\nu(\nu+1)+r(\nu+1)-\nu]}-q^{(j\nu+r)[j(\nu+1)+1]})\;;
\;\;\;\;\; r=1,2,\ldots,\nu-1,
\label{i2}
\end{equation}

\noindent
where for $i=1,2,\ldots,\nu-2$ and $m_i\in Z\geq 0$
\begin{equation}
\begin{array}{ccc}
(\vec{m})_i=m_i\;; & b^+=\nu-r-1\;; & b^-=r-1
\end{array}
\label{i3}
\end{equation}

\noindent
and $(\nu-2)\times(\nu-2)$-dimensional Cartan matrix $C$ is related to the
incidence matrix $K_{(\nu-2)}$
\begin{equation}
K_{\nu-2}= \left( \begin{array}{lcr}
010 & \cdots & 0 \\
101 & \cdots & 0 \\
\cdots & \cdots & \cdots \\
0 & \cdots & 101 \\
0 & \cdots & 010
\end{array} \right)
\label{i4}
\end{equation}

\noindent
in a standard fashion
\begin{equation}
C_{\nu-2}=2-K_{\nu-2}.
\label{i5}
\end{equation}

\noindent
The rest of notations in (\ref{i2}) will be defined later
(\ref{1.20}, \ref{1.21}, \ref{2.4}). The r.h.s. in (\ref{i2}) can be easily
recognized as Rocha-Caridi formula \cite{14} for the
$M(\nu,\nu+1)$ character
$(\chi_{r,s=1}^{\nu,\nu+1}(q)\sim Tr_{r,s=1}(q^{L_0}))$
of a primary field with conformal dimension $\Delta_{r,s=1}$
\begin{equation}
\Delta_{r,s}=\frac{[r(\nu+1)-s\cdot\nu]^2-1}{4\nu(\nu+1)}\;;
\;\;\;\;\; r=1,2,\ldots,\nu-1 ;\;\;\;\;\; s=1,2,\ldots,\nu
\label{i6}
\end{equation}

\noindent
and the l.h.s. in (\ref{i2}) stands for the companion fermionic representation,
proposed by the Stony Brook group \cite{3}.
Note that in the simplest case $\nu=3$ (Ising model) identity
(\ref{i2}) was known to be true for quite some time \cite{15}.
The first important step towards general proof was taken in \cite{6}
where polynomial generalization of (\ref{i2}) was proposed (see (\ref{2.54}))
and $\nu=4$ (and $3$) case was proven. The object of the present paper
is to provide proof for the remaining cases: $4<\nu<\infty$. \\
Let us now introduce some important background. Among many techniques
available to prove identities of the Rogers-Ramanujan type (\cite{15.5}
and references there), proof by Schur \footnote[1]{Since Schur
discovered and proved identities (\ref{i1}) independently,
it would be appropriate to correct the historical injustice
and call (\ref{i1}) Rogers-Ramanujan-Schur identities, as
attempted by some.}  \cite{16} still stands out
as, perhaps, a monument to the last days of the Kaiser Reich. The main
idea of Schur's approach (see also \cite{16.5}) was to convert $q$-series in
(\ref{i1}) into polynomials by introducing a finitization parameter $L$, which
roughly measures the degree of polynomials. Letting $L$ tend to
infinity, one recovers original identities. An advantage
of having $L$ is that it can be employed to prove polynomial
identities by means of recursion relations. Remarkably, this finitization
parameter has a direct physical interpretation. Depending on a
situation, one can think of $L$ either as a size of a system,
number of particles or as an ultraviolet cut-off for the truncated
conformal basis (\cite{6}, and below). \\
Another hero of our story is a $\nu$-state $RSOS$-model
introduced by Andrews,
Baxter, and Forrester (ABF) \cite{17} about ten years ago. Unlike some
papers which go through a long latent period before (if ever)
they are in the limelight, the ABF paper received immediate attention
due to Huse's identification \cite{18} of the critical points of
$RSOS$-models (in a first critical regime) with those described by
the unitary minimal series $M(\nu,\nu+1)$. Huse's observation was
later confirmed in \cite{19} and \cite{20}, where central charge
$c$ for a second critical regime of a $\nu$-state
$RSOS$-model was found to be equal to that of the Fateev-Zamolodchikov
parafermion model \cite{21}
$$ c=2-\frac{6}{\nu+1}. $$
The states of an $RSOS$-model can be thought as paths, labeled by
sequences of integers $\{a_0,a_1,\ldots,a_L\}$ with
$1\leq a_i\leq\nu$ $(i=0,1,\ldots,L)$, $a_0\equiv s, a_L\equiv r$.
In what follows, we will refer to the fixed values $r$ and $s$ as
boundary conditions $(r,s)$. \\
Boltzmann weights of a $\nu$-state $RSOS$-model (their explicit form
will not be needed here)
$$ W_{\nu}\left[\begin{array}{ccc}
    &  a'_{i+1} & \\
a_i &           & a_{i+2} \\
    &  a_{i+1}  &         \end{array}\right] $$
vanish, unless integers $a_i$'s satisfy the constraint
\begin{equation}
\mid a_i-a_{i+1}\mid =1.
\label{i7}
\end{equation}

\noindent
Dimension of the path space (subject to restriction (\ref{i7})
above and boundary conditions $(r,s)$) can be concisely expressed
in terms of the incidence matrix $K_{\nu}\;$ \footnote[2]{Throughout
this paper, subindex of the incidence matrix refers to its dimension.}
as
\begin{equation}
Dim_{r,s}(path\;space)=((K_{\nu})^L)_{s,r}.
\label{i8}
\end{equation}

\noindent
In the limit $L\rightarrow\infty$, it is possible to think of the
path states with boundary conditions $(s,r)$ as $M(\nu,\nu+1)$
Virasoro states in the module of a primary field with
conformal dimension $\Delta_{r,s}$ (\ref{i6}). \\
The corner transfer matrix calculations of ABF \cite{17} suggest
that the action of a Virasoro (energy) operator $L_0$ in the
path space $(r,s)$ can be described as
$$\frac{1}{4}\sum_{j=1}^{\infty}j\mid a_j-a_{j+2}\mid . $$
Not much is known about the action of other Virasoro generators
in the path space. The interested reader may consult \cite{22}-\cite{23.5}
for the recent developments.
Path-Virasoro isomorphism, once rigorously established, would
imply the natural interpretation of
finite $L$ as a kind of an ultraviolet cut-off for the truncated
conformal basis. \\
A connection between critical $RSOS$-model and $XXZ$-model with
rational anisotropy $\gamma=\frac{\pi}{\nu+1}$ was noticed in
\cite{20}, \cite{24} and \cite{24.5}, where it was shown that the spectrum of
an $RSOS$-model can be obtained from that of $XXZ$-model,
provided one implements certain projection mechanism. Moreover,
Reshetikhin proposed in \cite{25} that the $RSOS$-Boltzmann
weights appear as factors in the $S$-matrix, describing
scattering among the elementary excitations of spin $(\frac{\nu-1}{2})$
$XXZ$-model (in a weak anisotropic regime). According to \cite{25},
two-particle $S$-matrix has the following structure
\begin{equation}
S=W_{\nu}\otimes S_{SG},
\label{i9}
\end{equation}

\noindent
where the second factor in the r.h.s. of (\ref{i9}) denotes a
two-particle amplitude for the Sine-Gordon model. \\
We remind the reader that a fundamental particle of spin
$(\frac{\nu-1}{2}) XXZ$-model is associated with a hole in the
Dirac sea of negative energy Bethe strings of length $(\nu-1)$.
Other Bethe strings have zero energy. These strings can be used
to label internal quantum numbers of a fundamental particle.
In deriving (\ref{i9}), Reshetikhin identified short strings
(of length $1\leq l_i\leq\nu-2$) with $RSOS$ degrees of freedom
and long ones $(l_i\geq\nu)$ with vertex (spin $\frac{1}{2}$)
degrees of freedom of a particle. \\
In \cite{26} we used the Bethe ansatz technique to evaluate the dimension
of an "$RSOS$-string" space directly. Our calculations led us to
the following system of equations for the non-negative Bethe integers
$n_i,m_i(i=1,2,\ldots,\nu-2)$
\begin{eqnarray}
\left\{\begin{array}{l}
n_1+m_1=\frac{1}{2}m_2 \\
\cdots\cdots\cdots\cdots\cdots\cdots\cdots\cdots\cdot \\
n_i+m_i=\frac{1}{2}(m_{i+1}+m_{i-1}) \\
\cdots\cdots\cdots\cdots\cdots\cdots\cdots\cdots\cdot \\
n_{\nu-2}+m_{\nu-2}=\frac{1}{2}(L+m_{\nu-3}) \end{array}\right.
\label{i10}
\end{eqnarray}

\noindent
with $L$, now being a number of physical excitations. Depending on
$\nu$-parity (integer or half-integer spin), $L$ can be either
$$ L=0,1,2,\ldots\;;\;\;\;\;\;\nu\equiv even$$ or
$$ L=0,2,4,\ldots\;;\;\;\;\;\;\nu\equiv odd. $$
System (\ref{i10}) describes the $(\nu-2)$ Fermi bands. Each band
consists of the $(m_i+n_i)$ consecutive integers with only $n_i$
\underline{distinct} integers being "occupied". Because of this
exclusion rule, we would refer to (\ref{i10}) as a fermionic system. \\
It is a simple matter to convert (\ref{i10}) into a partition
problem for $\frac{L}{2}$
\begin{equation}
\sum_{i=1}^{\nu-2}n_i(\nu-1-i)+m_1\frac{\nu-1}{2}=\frac{L}{2}.
\label{i11}
\end{equation}

\noindent
To any set $\{m_1,n_1,n_2,\ldots,n_{\nu-2}\}$ satisfying (\ref{i11})
one adds a companion set
$\{m_2,m_3,\ldots,m_{\nu-2}\}$
$$m_i=2\sum_{j=1}^i(i-j)\cdot n_j +i\cdot m_1\;;\;\;\;\;\;
i=2,\ldots,\nu-2$$ to obtain all solutions to system (\ref{i10}). \\
Equation (\ref{i11}) clearly shows that it is impossible to
remove any "occupied" integer from a given Fermi band without
affecting the rest of them in a profound way. Thus, the fermionic
system (\ref{i10}) provides an interesting example of a
kinematic interaction. \\
At a first glance, path space dimension (\ref{i8}) and equations
(\ref{i10}) have nothing in common. However, motivated by
the extensive numerical checks, we conjectured in \cite{26} that
\footnote[3]{Throughout this paper $\left[\begin{array}{c} N \\ m
\end{array}\right]$ denotes the {\em usual} binomial coefficient.}
\begin{eqnarray}
((K_{\nu})^L)_{1,1}=\sum_{m_1\equiv even}\prod_{i=1}^{\nu-2}
\left[\begin{array}{c} n_i+m_i \\ n_i \end{array}\right]\;;
\;\;\;\;\; L\equiv even
\label{i12}
\end{eqnarray}
\begin{eqnarray}
((K_{\nu})^L)_{1,\nu}=\sum_{m_1\equiv odd}\prod_{i=1}^{\nu-2}
\left[\begin{array}{c} n_i+m_i \\ n_i \end{array}\right]\;;
\;\;\;\;\; L+\nu\equiv odd
\label{i13}
\end{eqnarray}

\noindent
with sum taken over the subset of all solutions to (\ref{i11})
which obey extra parity restrictions. \\
The original motivation for this paper was to provide
proof for the conjecture above.
While searching for proof, I became acquainted with
results of the Stony Brook group and got convinced that
fermionic description of the CFT characters is related to a
$q$-deformation of the counting formulas (\ref{i12}, \ref{i13}).
The remainder of this paper is organized as follows.\\
In Section 2, we generalize the fermionic system (\ref{i10})
in order to treat a more general class of boundary conditions.
Expressing generating functions for a number of paths and
fermionic states in terms of Chebyshev polynomials of the
second kind, we establish the equivalence of two counting
procedures.
In Section 3, we first, propose and prove a $q$-generalization of
certain binomial identities derived in Section 2 and then,
obtain the proof of fermionic sum representation for
$M(\nu,\nu+1)$-characters.
In Section 4, we discuss the physical significance of
our results and speculate about a possible generalization.
I conclude on a personal note.

\section*{2. Equivalence of fermionic and path counting of $RSOS$-states}

\setcounter{chapter}{2}
\setcounter{section}{2}
\setcounter{equation}{0}

In this section we find a closed form expression for generating
functions of number of path and fermionic states and, as a result,
establish the equivalence of different counting procedures.
For the sake of clarity, some technical details are relegated to
{\em Appendix A}.\\
The incidence matrix $K_{\nu}$ introduced in (\ref{i4})
has a well-known spectral decomposition
\begin{eqnarray}
K_{\nu}v^{(n)} & = & (2\cos\frac{\pi n}{\nu+1})v^{(n)}\;\;\;\;\;\;;
\;\;\;\;\; n=1,2,\ldots,\nu \nonumber \\
v_i^{(n)} & = & \sqrt{\frac{2}{\nu+1}}\cdot\sin\frac{\pi ni}{\nu+1}\;;
\;\;\;\;\;\; i=1,2,\ldots,\nu .
\label{1.1}
\end{eqnarray}

\noindent
We comment that it became a habit for some to refer to the parameters
$n$ and $(\nu+1)$ in formula (\ref{1.1}) as Coxeter exponents
and Coxeter number, respectively.
Keeping in mind that $L+r\equiv odd$ positive integer we introduce a
generating function
\begin{equation}
G_r^{(1)}(x)=\sum^{\infty}_{\stackrel{L=0}{L\equiv p_+(r)[mod 2]}}
x^{\frac{L}{2}}((K_{\nu})^L)_{1,r}=\left(\frac{1+p_+(r)(\sqrt xK_{\nu}-1)}
{1-xK_{\nu}^2}\right)_{1,r},
\label{1.2}
\end{equation}

\noindent
where $p_{\pm} (r)=\frac{1\pm (-1)^r}{2}$. Making use of a spectral
decomposition of the matrix $K_{\nu}$ (\ref{1.1}), we find
\begin{equation}
G_r^{(1)}(x)=\frac{2}{\nu+1}\sum_{n=1}^{\nu}\frac{1+p_+(r)
(2\sqrt x\cos\frac{\pi n}{\nu+1}-1)}{1-4x\cos^2\frac{\pi n}{\nu+1}}
\sin\frac{\pi nr}{\nu+1}\sin\frac{\pi n}{\nu+1}.
\label{1.3}
\end{equation}

\noindent
Introducing $z$ by
\begin{equation}
\frac{1}{\sqrt x}=z+\frac{1}{z}
\label{1.4}
\end{equation}

\noindent
we can present $G_r^{(1)}$ as
\begin{equation}
G_r^{(1)}(z)=\frac{2}{\nu+1}(z^2+1)^{p_{-}(r)+1}z^{p_{+}(r)}\sum_{n=1}^{\nu}
\frac{\sin(\frac{\pi nr}{\nu+1})\cdot\sin(\frac{\pi n}{\nu+1}(p_{+}(r)+1))}
{[z^2-exp(\frac{2\pi ni}{\nu+1})][z^2-exp(-\frac{2\pi ni}{\nu+1})]} .
\label{1.5}
\end{equation}

\noindent
To proceed further we need the formula below
\begin{equation}
(z^2+1)^{p_{-}(r)}(z^{2(\nu+1)}-1)\sum_{n=1}^{\nu}
\frac{\sin(\frac{\pi nr}{\nu+1})\cdot\sin(\frac{\pi n}{\nu+1}(p_{+}(r)+1))}
{[z^2-exp(\frac{2\pi ni}{\nu+1})][z^2-exp(-\frac{2\pi ni}{\nu+1})]}
=\frac{\nu+1}{2}(z^{2(\nu+1-r)}-1)z^{r-p_{+}(r)-1}.
\label{1.6}
\end{equation}

\noindent
The simplest way to prove (\ref{1.6}) is as follows. First, we check that
the l.h.s. of (\ref{1.6}), considered as function of $z^2$, has no poles.
This being so, we infer that both sides in (\ref{1.6}) are polynomials
of degree not higher than $\nu$. To complete the proof we evaluate
both sides at
\begin{equation}
z^2=exp(i\frac{2\pi}{\nu+1}j)\;;\;\;\;\;\;\;\;\;\;\;
j=0,1,2,\ldots,\nu .
\label{1.7}
\end{equation}

\noindent
After elementary (if tedious) calculations we see that identity (\ref{1.6})
holds for $z^2$, given by (\ref{1.7}). According to the fundamental
theorem of algebra, two polynomials which appear in (\ref{1.6}) must be equal.
Thus, formula (\ref{1.6}) is proven. \\
With help of (\ref{1.6}) we can process $G_r^{(1)}$ further
\begin{equation}
G_r^{(1)}=(z+\frac{1}{z})\times\frac{z^{\nu+1-r}-\frac{1}{z^{\nu+1-r}}}
{z^{\nu+1}-\frac{1}{z^{\nu+1}}} .
\label{1.8}
\end{equation}

\noindent
Introducing $\theta$
\begin{equation}
\frac{1}{\sqrt x}=z+\frac{1}{z}=2\cos \theta ,
\label{1.9}
\end{equation}

\noindent
one can express $G_r^{(1)}$ in terms of Chebyshev polynomials of the
second kind $U_m(\theta)$
\begin{equation}
G_r^{(1)}(\theta)=U_1(\theta)\cdot\frac{U_{\nu-r}(\theta)}{U_{\nu}(\theta)}\;;
\;\;\;\;\;\;\;\;\;\; U_m(\theta)=\frac{\sin(m+1)\theta}{\sin \theta} .
\label{1.10}
\end{equation}

\noindent
It is well known that orthogonal polynomials $U_m(\theta)$ satisfy
recurrences
\begin{equation}
2\cos\theta\cdot U_m(\theta)=U_{m-1}(\theta)+U_{m+1}(\theta)
\label{1.11}
\end{equation}

\noindent
along with initial conditions
\begin{equation}
U_0(\theta)=1\;;\;\;\;\;\; U_1(\theta)=2\cos\theta=\frac{1}{\sqrt x}.
\label{1.12}
\end{equation}

\noindent
There is, however, another way to specify these polynomials uniquely
\begin{eqnarray}
U_{m-1}(\theta)\cdot U_{m+1}(\theta)=U_m^2(\theta)-1 \nonumber \\
\;\;U_0(\theta)=1\;;\;\;\;\;\;\;\;\; U_1(\theta)=2\cos\theta .
\label{1.13}
\end{eqnarray}

\noindent
For a reason which will become apparent later, we'd like to call
(\ref{1.13}) fermionic recurrences. \\
As immediate consequences of results (\ref{1.8}, \ref{1.10}, \ref{1.11})
we have for $L+r\equiv odd$
\begin{eqnarray}
((K_{\nu})^L)_{1,r} & = & \frac{1}{2\pi i}
\oint\frac{dx}{x}\frac{G_r^{(1)}}{x^{L/2}}
=\frac{1}{2\pi i}\oint\frac{dz}{z}(z+\frac{1}{z})^Lz^{1-r}(1-z^{-2})
\frac{1-z^{-2(\nu+1-r)}}{1-z^{-2(\nu+1)}}=   \nonumber \\[3mm]
& = & \sum_{j=-\infty}^{\infty}\left\{ \left[ \begin{array}{c}
L \\ \frac{L+1-r}{2}-j(\nu+1) \end{array} \right] - \left[ \begin{array}{c}
L \\ \frac{L-1-r}{2}-j(\nu+1) \end{array} \right] \right\}
\label{1.14}
\end{eqnarray}
and
\begin{eqnarray}
((K_{\nu})^L)_{1,r+1}+((K_{\nu})^L)_{1,r-1}
& = & \frac{1}{2\pi i}\oint\frac{dx}{x}\frac{1}{x^{L/2}}\frac{U_1}{U_{\nu}}
(U_{\nu -r-1}+U_{\nu -r+1})=  \nonumber \\[2mm]
& = & \frac{1}{2\pi i}\oint\frac{dx}{x}\frac{1}{x^{\frac{L+1}{2}}}
\frac{U_1}{U_{\nu}}U_{\nu-r}=((K_{\nu})^{L+1})_{1,r}.
\label{1.15}
\end{eqnarray}

\noindent
In what follows, we would frequently refer to the r.h.s. of (\ref{1.14}) as
bosonic counting of the path space.
The last equation clearly exhibits a remarkable connection between recurrences
for path counting function $((K_{\nu})^L)_{1,r}$ and that for the orthogonal
polynomials (\ref{1.11}).
We intend to pursue this observation further elsewhere. \\
We now move on to introduce a different set of definitions, related to what we
call fermionic counting of the $RSOS$ states.  First, we generalize
system (\ref{i10}) by adding to it the inhomogeneous term, parametrized by
$r=1,2,\ldots,\nu-p_{+}(L+\nu)$. For $i=1,2,\ldots,\nu-2$ let it be written
\begin{eqnarray}
n_i+m_i & = & \frac{1}{2}\left\{ \theta(i>1)\cdot m_{i-1}
+\theta(\nu-2>i)\cdot m_{i+1}+
L\cdot\delta_{i,\nu-2} \right\} \nonumber  \\
& + & \frac{1}{2} \left\{
p_{-}(L+r)\cdot\delta_{i,r-1}+p_{+}(L+r)
\cdot\delta_{i,\nu-1-r}\right\} \nonumber \\[2mm]
m_1 & = & 2n_0+p_{-}(L+r)\cdot\theta(r>1)-p_+(L+r)\cdot\delta_{r,\nu-1}
\label{1.16}
\end{eqnarray}

\noindent
where $n_0, n_1,\cdots,n_{\nu-2}$ are non-negative integers and
$\delta_{i,j},\theta(i>j)$ denote Kronecker delta and a step function,
respectively. Again,
it is trivial to show that system (\ref{1.16}) is equivalent to the
following partition problem
\begin{equation}
\sum_{i=0}^{\nu-2}n_i(\nu-1-i)=
\frac{L+r\cdot p_{+}(L+r)-(r-1)\cdot p_{-}(L+r)}{2}
\;;\;\;\;\;n_i\geq 0 \mbox{ for } i=0,1,\ldots,\nu-2.
\label{1.17}
\end{equation}

\noindent
Once a set of non-negative integers $\{n_i\}$ which satisfies constraint
(\ref{1.17}) is found, set $\{m_i\}$ can be determined for
$i=1,2,\ldots,\nu-2$ as
\begin{equation}
m_i = \tilde{n}_i-V_{i,r}p_{+}(L+r)+\tilde{V}_{i,r}p_{-}(L+r)
\label{1.18}
\end{equation}
where
\begin{equation}
\tilde{n}_i = 2\sum_{l=1}^i ln_{i-l}
\label{1.19}
\end{equation}
\begin{equation}
V_{i,r}\equiv V_{i,r}^+=[i-(\nu-1-r)]\cdot\theta(i>\nu-1-r)
\label{1.20}
\end{equation}
\begin{equation}
\tilde{V}_{i,r}\equiv V_{i,r}^-=(r-1)+[i-(r-1)]\cdot\theta(r-1>i)
\label{1.21}
\end{equation}

\noindent
Next we define fermionic counting functions $I(L,r)$ and $J(L,r)$
\begin{eqnarray}
I(L,r) = \sum\prod_{i=1}^{\nu-2}\left[\begin{array}{c}
n_i+m_i \\ n_i \end{array} \right]=\sum\prod_{i=1}^{\nu-2}
\left[\begin{array}{c}
n_i+\tilde{n}_i+\tilde{V}_{i,r} \\ n_i \end{array} \right] ; \;\;\;\;\;
\begin{array}{l}
L+r\equiv odd \\ 1\leq r\leq\nu \end{array}
\label{1.22}
\end{eqnarray}

\noindent
\begin{eqnarray}
J(L,r) = \sum\prod_{i=1}^{\nu-2}\left[\begin{array}{c}
n_i+m_i \\ n_i \end{array} \right]=\sum\prod_{i=1}^{\nu-2}
\left[\begin{array}{c}
n_i+\tilde{n}_i-V_{i,r} \\ n_i \end{array} \right] ; \;\;\;\;\;
\begin{array}{l}
L+r\equiv even \\ 1\leq r\leq\nu-1 \end{array}
\label{1.23}
\end{eqnarray}

\noindent
where symbol $\sum$ above stands for the sum over all solutions to
constraint (\ref{1.17}). We'd like to point out that even though we did not
impose any constraints on $\{m_i\}$, except (\ref{1.18}), it is clear
from definition (\ref{1.22}, \ref{1.23}) that effectively
$m_i\geq 0$, since for
$m_i<0$ corresponding binomial coefficient $\left[ \begin{array}{c}
n_i+m_i \\ m_i \end{array}\right]$ vanishes. By the same argument, one may
disregard requirement $n_i\geq 0$. \\
We are now in a position to formulate the main assertion of this section
\begin{equation}
((K_{\nu})^L)_{1,r} = I(L,r) = J(L,r-1) \;;\;\;\;\;\;\;\;\;L+r\equiv odd.
\label{1.24}
\end{equation}

\noindent
For the proof, let's construct two fermionic generating functions
\begin{equation}
G_r^{(2)}(x)=\sum^{\infty}_{\stackrel{L=0}{L\equiv p_{+}(r)(mod2)}}
x^{\frac{L}{2}}
I(L,r)\;;\;\;\;\;\;\;\;\;\;\;L+r\equiv odd
\label{1.25}
\end{equation}
and
\begin{equation}
G_r^{(3)}(x)=\sum^{\infty}_{\stackrel{L=0}{L\equiv p_{-}(r)(mod2)}}
x^{\frac{L}{2}}
J(L,r)\;;\;\;\;\;\;\;\;\;\;\;L+r\equiv even .
\label{1.26}
\end{equation}

\noindent
Taking into account constraint (\ref{1.17}), one finds
\begin{equation}
G_r^{(2)}(x)=\sum^{\infty}_{n_0,n_1,\ldots,n_{\nu-2}=0} x^{\frac{r-1}{2}+
\sum_{i=0}^{\nu-2}n_i(\nu-1-i)}\times
\prod_{i=1}^{\nu-2}\left[\begin{array}{c}
n_i+\tilde{n}_i+\tilde{V}_{i,r} \\ n_i \end{array} \right] .
\label{1.27}
\end{equation}

\noindent
To proceed further we shall recall the binomial theorem
\begin{equation}
\sum_{n_i=0}^{\infty}\left[\begin{array}{c}
n_i+\tilde{n}_i+\tilde{V}_{i,r} \\ n_i \end{array} \right]\tilde x^{n_i}=
\frac{1}{(1-\tilde x)^{\tilde n_i+1+\tilde{V}_i,r}}.
\label{1.28}
\end{equation}

\noindent
Our strategy for obtaining closed form expression for $G_r^{(2)}$ is to apply
(\ref{1.28}) in repetitive fashion to sum out all variables $n_i$'s.
Let's demonstrate how it works. Having summed out $n_{\nu-2}$-variable in
(\ref{1.27}) we obtain with help of $\tilde{n}_{\nu-2}=\sum_{l=0}^{\nu-3}
(\nu-2-l)n_l$
\begin{equation}
G_r^{(2)}=\sum^{\infty}_{n_0,n_1,\ldots,n_{\nu-3}=0} x^{\frac{r-1}{2}}
\frac{1}{(1-T_{1,0})^{1+\tilde{V}_{\nu-2,r}}}\times
\prod_{i=0}^{\nu-3}(T_{\nu-1-i,1})^{n_i}\times
\prod_{i=1}^{\nu-3}\left[\begin{array}{c}
n_i+\tilde{n}_i+\tilde{V}_{i,r} \\ n_i \end{array} \right],
\label{1.29}
\end{equation}
where
\begin{equation}
T_{i,1}=\frac{T_{i,0}}{(1-T_{1,0})^{2(i-1)}}
\label{1.30}
\end{equation}
and
\begin{equation}
T_{i,0}=x^i\;;\;\;\;\;\;\;\;\;\;\;\;\;\;\;\;i=1,2,\ldots.
\label{1.31}
\end{equation}

\noindent
Obviously, one can use the binomial theorem (\ref{1.28}) again to eliminate
$n_{\nu-3}$-variable.
\begin{eqnarray}
G_r^{(2)} & = & \sum^{\infty}_{n_0,n_1,\ldots,n_{\nu-4}=0}
x^{\frac{r-1}{2}}\cdot
\frac{1}{(1-T_{1,0})^{1+\tilde{V}_{\nu-2,r}}}\cdot
\frac{1}{(1-T_{2,1})^{1+\tilde{V}_{\nu-3,r}}}\times \nonumber \\[3mm]
& \times & \prod_{i=0}^{\nu-4}(T_{\nu-1-i,2})^{n_i}\times
\prod_{i=1}^{\nu-4}\left[\begin{array}{c}
n_i+\tilde{n}_i+\tilde{V}_{i,r} \\ n_i \end{array} \right] ,
\label{1.32}
\end{eqnarray}
where
\begin{equation}
T_{i,2}=\frac{T_{i,1}}{(1-T_{2,1})^{2(i-2)}}.
\label{1.33}
\end{equation}

\noindent
Proceeding as above, we derive
\begin{equation}
G_r^{(2)}=x^{\frac{r-1}{2}}\prod_{i=1}^{\nu-1}
\frac{1}{(1-T_{i,i-1})^{1+\tilde{V}_{\nu-1-i,r}}}
\label{1.34}
\end{equation}

\noindent
with quantities $T_{i,m}$ being determined recursively as
\begin{eqnarray}
\left\{ \begin{array}{lll}
T_{i,m}=\frac{T_{i,m-1}}{(1-T_{m,m-1})^{2(i-m)}} & ; &  m=0,1,2,\ldots \\[2mm]
T_{i,0}=x^i & ; & i=1,2,\ldots .
\end{array}\right.
\label{1.35}
\end{eqnarray}

\noindent
The reader is encouraged to prove (\ref{1.35}) by carrying out
the inductive step $m\rightarrow m+1$. \\
As shown in {\em Appendix A}, recurrences (\ref{1.35}) imply interesting
connection between $T_{j,m}$'s and Chebyshev polynomials $U_m$'s.
In particular,
\begin{equation}
T_{m,m-1}=U_m^{-2}(\theta).
\label{1.36}
\end{equation}

\noindent
Now we substitute (\ref{1.21}) and (\ref{1.36}) into (\ref{1.34}) to get
\begin{equation}
G_r^{(2)}=x^{\frac{r-1}{2}}\cdot\prod_{m=1}^{\nu-r} \left(
\frac{U^2_m}{U_m^2-1} \right)^r
\times\prod_{m=\nu-r+1}^{\nu-1}\left(\frac{U^2_m}{U_m^2-1}\right)^{\nu-m}.
\label{1.37}
\end{equation}

\noindent
Using fermionic recurrences (\ref{1.13}) for $U_m$, two products
above telescope to yield a final result
\begin{equation}
G_r^{(2)}=x^{\frac{r-1}{2}}\cdot
\left(\frac{U_{\nu-r}U_1}{U_{\nu-r+1}}\right)^r
\cdot\left[\left(\frac{U_{\nu-r+1}}{U_{\nu-r}}\right)^r
\cdot\frac{U_{\nu-r}}{U_{\nu}}
\right] =U_1\cdot\frac{U_{\nu-r}}{U_{\nu}}.
\label{1.38}
\end{equation}

\noindent
Comparing (\ref{1.10}) and (\ref{1.38}) it is plain
\begin{equation}
G_r^{(1)}=G_r^{(2)}
\label{1.39}
\end{equation}
and therefore,
\begin{equation}
((K_{\nu})^L)_{1,r}=I(L,r).
\label{1.40}
\end{equation}

\noindent
Treating $G_r^{(3)}$ in a similar fashion we first find
\begin{equation}
G_r^{(3)}=x^{-\frac{r}{2}}\cdot\prod_{m=1}^{r-1}(1-U_m^{-2})^{r-1-m}\cdot
\prod_{m'=r}^{\nu-1}\frac{1}{(1-U_{m'}^{-2})}
\label{1.41}
\end{equation}

\noindent
and then, use fermionic recurrences (\ref{1.13}) again to arrive
at a simple form
\begin{equation}
G_r^{(3)}=\frac{U_1\cdot U_r\cdot U_{\nu-1}}{U_{\nu}}=
G_{r+1}^{(2)}+U_1\cdot U_{r-1}.
\label{1.42}
\end{equation}
Since
\begin{equation}
\oint\frac{dx}{x}\cdot\frac{U_1\cdot U_{r-1}}{x^{\frac{L}{2}}}=0 \;;
\;\;\;\;\;\;\;\;\;\;L=0,1,2,\ldots
\label{1.43}
\end{equation}

\noindent
we immediately have
\begin{equation}
((K_{\nu})^L)_{1,r}=I(L,r)=\frac{1}{2\pi i}\cdot\oint\frac{dx}{x}
\frac{G_r^{(2)}}{x^{\frac{L}{2}}}=\frac{1}{2\pi i}\cdot\oint\frac{dx}{x}
\frac{G_{r-1}^{(3)}}{x^{\frac{L}{2}}}=J(L,r-1).
\label{1.44}
\end{equation}

\noindent
That concludes our proof of (\ref{1.24}). \\
At this point it is natural to combine $I(L,r)$ and $J(L,r)$ into a single
fermionic object $F(L,r)$
\begin{eqnarray}
F(L,r)=\left\{\begin{array}{cl}
I(L,r)\;; & \;\;\;\;\;\;\;\;\;\;L+r\equiv odd \\
J(L,r)\;; & \;\;\;\;\;\;\;\;\;\;L+r\equiv even .
\end{array}\right.
\label{1.45}
\end{eqnarray}

\noindent
We shall now use identities (\ref{1.14}) and (\ref{1.24}) to express the
main result of this section in a following form
\[
F(L,r)=\sum\prod_{i=1}^{\nu-2}\left[\begin{array}{c}
n_i+m_i \\ n_i \end{array}\right]= \]
\begin{eqnarray}
\sum_{j=-\infty}^{\infty}
\left\{ \left[\begin{array}{c}
L \\ ((\frac{L+1-r}{2}))-j(\nu+1) \end{array}\right]
- \left[\begin{array}{c}
L \\ ((\frac{L-1-r}{2}))-j(\nu+1) \end{array}\right] \right\},
\label{1.46}
\end{eqnarray}

\noindent
where $((x))$ denotes the integer part of $x$. Introducing compact vector
notations
\begin{equation}
(\vec{m})_i = m_i
\label{1.47}
\end{equation}
\begin{equation}
(\vec{V}_r(L))_i = -p_{+}(L+r)\cdot V_{i,r}+p_{-}(L+r)
\cdot\tilde{V}_{i,r}
\label{1.48}
\end{equation}
\begin{equation}
(\vec{u}_r(L))_i = p_{+}(L+r)\cdot\delta_{i,\nu-r-1}+p_{-}(L+r)
\cdot\delta_{i,r-1}
\label{1.49}
\end{equation}

\noindent
and remembering the system of equations (\ref{1.16}), we can rewrite the
fermionic sum in (\ref{1.46}) entirely in terms of
$\vec{m}$-variables.
\[
\sum_{\vec{V}_r(L)}\cdot\prod_{i=1}^{\nu-2}\left[\begin{array}{c}
\frac{1}{2}(K_{\nu-2}\cdot\vec{m}+\vec{u}_r(L))_i+
\frac{1}{2}L\delta_{i,\nu-2} \\ m_i \end{array}\right]= \]
\begin{eqnarray}
\sum_{j=-\infty}^{\infty}\left\{ \left[\begin{array}{c}
L \\ ((\frac{L+1-r}{2}))-j(\nu+1) \end{array}\right] -
\left[\begin{array}{c}
L \\ ((\frac{L-1-r}{2}))-j(\nu+1) \end{array}\right]\right\}
\label{1.50}
\end{eqnarray}

\noindent
where symbol $\sum_{\vec{V}_r(L)}$ stands for the sum over
\begin{equation}
m_i\equiv(\vec{V}_r(L))_i(mod 2)\;;\;\;\;\;\;m_i\in Z\;;\;\;\;\;\;
i=1,2,\ldots,\nu-2 .
\label{1.51}
\end{equation}

\noindent
The advantage of representation (\ref{1.50}) is that the summation
variables $\vec{m}$ are almost free, subject only to constraint
(\ref{1.51}). Also, as we shall see in the next section, a $q$-analogue
of representation (\ref{1.50}) is particularly well - suited to study
$L\rightarrow\infty$ limit. It should be added that identities similar to
(\ref{1.50}) for $\nu=5$ case ($3$-state Potts model) were proven in
\cite{27}\\
It follows from elementary recurrences
\begin{equation}
\left[ \begin{array}{c} N \\ m \end{array} \right] =
\left[ \begin{array}{c} N-1 \\ m-1 \end{array} \right] +
\left[ \begin{array}{c} N-1 \\ m \end{array} \right]
\label{1.52}
\end{equation}
and (\ref{1.46}) that for $1<r<\nu$
\begin{equation}
F(L,r)=F(L-1,r)+ \left\{\begin{array}{cl}
F(L-1,r-1)\;; & \;\;\;\;\;\;\;\;\;\;L+r\equiv odd \\
F(L-1,r+1)\;; & \;\;\;\;\;\;\;\;\;\;L+r\equiv even \\
\end{array}\right.
\label{1.53}
\end{equation}
or, equivalently
\begin{equation}
\left\{ \begin{array}{l}
J(L,r)=I(L-1,r)+J(L-1,r+1) \\
I(L,r)=J(L-1,r)+I(L-1,r-1). \end{array}\right.
\label{1.54}
\end{equation}

\noindent
Surely, since $I(L,r)=J(L,r-1)$, one may be tempted to rewrite the above
in a simpler form
\begin{equation}
\left\{ \begin{array}{l}
J(L,r)=J(L-1,r+1)+J(L-1,r-1) \\
I(L,r)=I(L-1,r+1)+I(L-1,r-1). \end{array}\right.
\label{1.55}
\end{equation}

\noindent
I believe that recurrences (\ref{1.54}) are more natural than (\ref{1.55}),
if one insists on interpreting $J(L,r)$ and $I(L,r)$ as components of a single
object $F(L,r)$.
As further evidence for this point of view, we mention (somewhat anticipating
things to come) that it is (\ref{1.54}) rather than (\ref{1.55}) which
admits straightforward $q$-generalization and that for $q\neq 1$,
identity
\begin{equation}
I(L,r)=J(L,r-1)
\label{1.56}
\end{equation}
no longer holds true. \\
Nevertheless, if one insists on casting (\ref{1.55}) in a form which
mimics $q$-case, then (\ref{1.55}) should be rewritten as
\begin{equation}
\left\{ \begin{array}{l}
\{J(L,r)-J(L-1,r+1)\}=J(L-2,r)+\{J(L-1,r-1)-J(L-2,r)\} \\
\{I(L,r)-I(L-1,r-1)\}=I(L-2,r)+\{I(L-1,r+1)-I(L-2,r)\}. \end{array} \right.
\label{1.57}
\end{equation}

\noindent
In this section we've demonstrated the equivalence of fermionic and
bosonic counting of the $RSOS$-states by means of a generating function
approach. As an alternative way of establishing this equivalence, one may
try to prove directly that $F(L,r)$ satisfies recurrences (\ref{1.53}).
This alternative becomes especially valuable since the author did not
(yet) succeed in finding a $q$-analogue of a generating function
method. \\
We now venture into the realm of $q$-identities.

\section*{3. Proof of the $q$-identities}

\setcounter{chapter}{3}
\setcounter{section}{0}
\setcounter{equation}{0}

We start this section by reminding the reader a few $q$-definitions. The
$q$-generalization of number $X$, due to Heine \cite{28}, can be written as
\begin{equation}
X_q=\frac{1-q^X}{1-q}.
\label{2.1}
\end{equation}
As $q$ tends to one
\begin{equation}
\lim_{q \rightarrow 1}X_q=X.
\label{2.2}
\end{equation}

\noindent
Next, we introduce $q$-shifted factorial $(a,q)_n$
\begin{equation}
(a,q)_n=\left\{ \begin{array}{cll}
1 & ; & \;\;\;\;\;n=0 \\[2mm]
\prod_{j=0}^{n-1}(1-aq^j) & ; & \;\;\;\;\;n\geq 1,
\end{array} \right.
\label{2.3}
\end{equation}

\noindent
where $(n\in Z)$ and (for $N$, $m\in Z$) $q$-binomial coefficient
\begin{equation}
\;\;\;\left[ \begin{array}{c}
N \\ m
\end{array} \right]_q
= \left\{ \begin{array}{cll}
\frac{(q,q)_N}{(q,q)_m(q,q)_{N-m}} & ; & \;\;\;\;\;0\leq m\leq N \\[2mm]
0 & ; & \;\;\;\;\;\mbox{otherwise.}
\end{array} \right.
\label{2.4}
\end{equation}

\noindent
It is quite amusing that $q$-numbers were rediscovered recently
in the context of quantum groups. There, symbol $q$ naturally refers to the
quantum deformation parameter. But Heine and others used the same $q$-symbol
almost a century before the advent of Quantum mechanics. What is it? Mere
coincidence or, maybe, Heine somehow knew about Quantum mechanics? If so, then
this is just another hint that the plane of knowledge exists outside of time
and space, in some ideal Plato's world of ideas \cite{29}. Returning
to our mundane business,
we note elementary recursion relations for the $q$-binomial coefficients
\begin{equation}
\left[ \begin{array}{c} N \\ m \end{array} \right]_q =
\left[ \begin{array}{c} N-1 \\ m-1 \end{array} \right]_q + q^m
\left[ \begin{array}{c} N-1 \\ m \end{array} \right]_q
\label{2.5}
\end{equation}

\begin{equation}
\;\;\;\left[ \begin{array}{c} N \\ m \end{array} \right]_q =
\left[ \begin{array}{c} N-1 \\ m \end{array} \right]_q + q^{N-m}
\left[ \begin{array}{c} N-1 \\ m-1 \end{array} \right]_q
\label{2.6}
\end{equation}

\noindent
which are similar to those for the usual binomial coefficients (\ref{1.52}).
On the other hand, there are certain limiting procedures which are not well
defined for $q=1$, but may be perfectly well defined
for $\mid q\mid <1$. For instance, letting $N$ tend to infinity, one finds
\begin{eqnarray}
\lim_{N\rightarrow \infty}
\left[ \begin{array}{c} N \\ m \end{array} \right]_q = \frac{1}{(q,q)_m} ;
& \;\;\;\;\;\;\;\;\;\;\mid q\mid <1
\label{2.7}
\end{eqnarray}
and
\begin{eqnarray}
\;\lim_{\stackrel{N\rightarrow\infty}{m\rightarrow\infty}}
\left[ \begin{array}{c} N \\ m \end{array} \right]_q =
\frac{1}{(q,q)_{\infty}} ;
& \;\;\;\;\;\;\;\;\;\;\mid q\mid <1,\;\;N>m .
\label{2.8}
\end{eqnarray}

\noindent
We are now well equipped to propose a $q$-analogue of the fermionic
counting functions (\ref{1.22}, \ref{1.23}).
For $r=1,2,\ldots,\nu$ let us introduce
\begin{eqnarray}
I_q(L,r)=\sum q^{Y_r(L)}\prod_{i=1}^{\nu-2}\left[\begin{array}{c}
n_i+m_i \\ n_i \end{array}\right]_q=
\sum q^{Y_r(L)}\prod_{i=1}^{\nu-2}\left[\begin{array}{c}
n_i+\tilde{n}_i+\tilde{V}_{i,r} \\ n_i \end{array}\right]_q ;
\;\;\; L+r\equiv odd
\label{2.9}
\end{eqnarray}

\noindent
and for $r=1,2,\ldots,\nu-1$
\begin{equation}
J_q(L,r)=\sum q^{X_r(L)}\prod_{i=1}^{\nu-2}\left[\begin{array}{c}
n_i+m_i \\ n_i \end{array}\right]_q=
\sum q^{X_r(L)}\prod_{i=1}^{\nu-2}\left[\begin{array}{c}
n_i+\tilde{n}_i-V_{i,r} \\ n_i \end{array}\right]_q ;
\;\; L+r\equiv even \;\;
\label{2.10}
\end{equation}
where
\begin{equation}
X_r(L)=-\frac{1}{2}\sum_{i=1}^{\nu-2}(n_i-\frac{L}{2}\delta_{i,\nu-2}-
\frac{1}{2}\delta_{i,\nu-r-1})\cdot(\tilde{n}_i-V_{i,r})
\label{2.11}
\end{equation}
and
\begin{equation}
Y_r(L)=-\frac{1}{2}\sum_{i=1}^{\nu-2}(n_i-\frac{L}{2}\delta_{i,\nu-2}-
\frac{1}{2}\delta_{i,r-1})\cdot(\tilde{n}_i+\tilde{V}_{i,r})
\label{2.12}
\end{equation}

\noindent
with the rest of notations the same as in {\em Section 2}. \\
{}From the definition above, it is obvious
\begin{equation}
I_q(L,\nu)=I_q(L-1,\nu-1)=J_q(L,\nu-1)
\label{2.13}
\end{equation}
and
\begin{equation}
J_q(L-1,1)=I_q(L,1).
\label{2.14}
\end{equation}

\noindent
What, perhaps, is not so obvious that $I_q$'s and $J_q$'s satisfy the
following recursion relations
\begin{eqnarray}
\{I_q(L,r)-\theta(r>1)\cdot
q^{\frac{L}{2}}I_q(L-1,r-1)\}=\;\;\;\;\;\;\;\;\;\;\;\;\;\;\;\;\;\;\nonumber \\
I_q(L-2,r)+\theta(\nu-1>r)\cdot q^{\frac{L-1}{2}}
\{I_q(L-1,r+1)-q^{\frac{L-1}{2}}\cdot I_q(L-2,r)\}
\label{2.15}
\end{eqnarray}
and
\begin{eqnarray}
\{J_q(L,r)-\theta(\nu-1>r)\cdot
q^{\frac{L}{2}}\cdot J_q(L-1,r+1)\}=\;\;\;\;\;\;\;\;\;\;\;\;  \nonumber \\
J_q(L-2,r)+\theta(r>1)\cdot q^{\frac{L-1}{2}}
\{J_q(L-1,r-1)-q^{\frac{L-1}{2}}\cdot J_q(L-2,r)\},
\label{2.16}
\end{eqnarray}

\noindent
with $\theta(i>j)$ being a step function
$$ \theta(i>j)\equiv \left\{ \begin{array}{ll}
1\;, & i>j \\ 0\;, & i\leq j . \end{array}\right. $$

\noindent
We'd like to comment here that from the technical point of view the choice
of phase factors (\ref{2.11}, \ref{2.12}) is custom-tailored to support
a proof of recurrences (\ref{2.15}, \ref{2.16}), which in turn were
motivated by $q=1$ case. Yet, from the physical point of view, the choice
(\ref{2.11}, \ref{2.12}) is intimately related to the assumption of linear
dispersion law for quasi-particles as we shall later see. \\
To set a stage for the technically involved proof of (\ref{2.15})
and (\ref{2.16}), in this section, we consider only (\ref{2.16}) with
$r=1,2$ and relegate our general discussion to {\em Appendices B} and
{\em C}. Before moving on, a few remarks are in order. In what follows,
the expression "telescopic expansion" will be frequently used.
Since it is hardly
a standard terminology we have to comment on its meaning. For
$a=0,1$ and $i=0,1,\ldots,n$ let $A_i^{(a)}$ be some
product of the $q$-binomial coefficients. We call the sum
$A_0^{(1)}+\sum_{i=1}^n A^{(0)}_i$ telescopic expansion for $A_n^{(1)}$
if
\begin{equation}
A_i^{(1)}+A^{(0)}_{i+1}=A_{i+1}^{(1)}
\label{2.17}
\end{equation}
holds.
Properties (\ref{2.17}), which are somewhat similar to that of the sliding
tubes of a jointed telescope, clearly imply
\begin{equation}
A_n^{(1)}=A_0^{(1)}+\sum_{i=1}^n A^{(0)}_i .
\label{2.18}
\end{equation}

\noindent
We note that recurrences (\ref{2.5}, \ref{2.6}) provide a simple example
of a telescopic expansion \footnote[4]{It is also possible to give a
"particle" interpretation to properties (\ref{2.17}, \ref{2.18}), however,
I prefer "telescopic" analogy because it is more visual}.\\
Our second remark concerns with interesting features of the solutions
to system (\ref{1.16}). To make notations more manageable we
introduce symbol $\{ \vec{n},\vec{\tilde n} \}_{L,r}$
which denotes some particular set of integers $(\vec{n})_i\equiv n_i$ and
$(\vec{\tilde n})_i\equiv\tilde n_i$, defined by (\ref{1.17})
and (\ref{1.19}). The following properties of
$\{\vec{n},\vec{\tilde n}\}_{L,r}$ sets are indispensable for
our treatment
\begin{eqnarray}
\{ \vec{n},\vec{\tilde n} \}_{L,r}
=\left\{ \begin{array}{l}
\{\vec{n},\vec{\tilde n}\}_{L-1,r+1}\;;
\;\;\;\;\;\;\;\;\;\; L+r\equiv even \\
\{\vec{n},\vec{\tilde n}\}_{L-1,r-1}\;;
\;\;\;\;\;\;\;\;\;\; L+r\equiv odd
\end{array} \right.
\label{2.19}
\end{eqnarray}
\begin{eqnarray}
\{\vec{n},\vec{\tilde n}\}_{L-2,r} = \{\vec{n},
\vec{\tilde n}\}_{L,r}-\{\vec{e}_{a=\nu-2},0 \}
\label{2.20}
\end{eqnarray}

\noindent
with unit vector $\vec{e}_a$, defined by its components as
\begin{equation}
(\vec{e}_a)_i=\delta_{i,a}\;;
\;\;\;\;\;\;\;\;\;\; a=1,2,\ldots,\nu-2 .
\label{2.21}
\end{equation}

\noindent
Confirmation of (\ref{2.19}) and (\ref{2.20}) is a simple
matter and we leave it as an exercise for the reader.\\
Let us now turn to the proof of (\ref{2.16}) for $r=1$, which is
an easiest possible case. \\
Using $q$-binomial identities (\ref{2.5}), we expand $J_q(L,1)$ as
\begin{eqnarray}
J_q(L,1) & = & \sum q^{X_1(L)}\prod_{i=1}^{\nu-2}\left[\begin{array}{c}
n_i-\delta_{i,\nu-2}+\tilde{n}_i \\ n_i-\delta_{i,\nu-2}
\end{array}\right]_q + \nonumber \\[3mm]
& + & \sum q^{X_1(L)+n_{\nu-2}}\prod_{i=1}^{\nu-2}\left[\begin{array}{c}
n_i+\tilde{n}_i-\delta_{i,\nu-2} \\ n_i \end{array}\right]_q .
\label{2.22}
\end{eqnarray}

\noindent
Keeping in mind (\ref{2.19}, \ref{2.20}), it is trivial to verify
\begin{eqnarray}
X_1(L) & = & X_1(L-2) \nonumber \\
X_1(L)+n_{\nu-2} & = & \frac{L}{2}+X_2(L-1) .
\label{2.23}
\end{eqnarray}

\noindent
Results (\ref{2.22}) and (\ref{2.23}), together with
\begin{equation}
\{\vec{n},\vec{\tilde n}\}_{L-2,1}=\{\vec{n},
\vec{\tilde n}\}_{L,1}-\{\vec{e}_{\nu-2},0\}
\label{2.24}
\end{equation}
\begin{equation}
\{\vec{n},\vec{\tilde n}\}_{L-1,2}=\{\vec{n},
\vec{\tilde n}\}_{L,1}\;\;\;\;\;\;\;\;\;\;
\label{2.25}
\end{equation}
\begin{equation}
V_{i,2} = \delta_{i,\nu-2}\;\;\;\;\;\;
\label{2.26}
\end{equation}

\noindent
allow us to recognize expansion (\ref{2.22}) for $J_q(L,1)$ as
\begin{equation}
J_q(L,1)=J_q(L-2,1)+q^{\frac{L}{2}}\cdot J_q(L-1,2).
\label{2.27}
\end{equation}

\noindent
This proves recurrences (\ref{2.16}) for $r=1$. \\
Treatment of $r=2$ case is more interesting,
since it involves a new element which plays an important role in a
general case. Once again, we construct a telescopic expansion for
$J_q(L,2)$
\begin{eqnarray}
J_q(L,2) & = & \sum q^{X_2(L)+n_{\nu-2}+n_{\nu-3}}
\prod_{i=1}^{\nu-2}\left[\begin{array}{c}
n_i+\tilde{n}_i-2\delta_{i,\nu-2}-\delta_{i,\nu-3} \\ n_i
\end{array}\right]_q  \nonumber \\[3mm]
& + & \sum q^{X_2(L)+n_{\nu-2}}\prod_{i=1}^{\nu-2}\left[\begin{array}{c}
n_i+\tilde{n}_i-2\delta_{i,\nu-2}-\delta_{i,\nu-3} \\ n_i-\delta_{i,\nu-3}
\end{array}\right]_q \nonumber \\[3mm]
& + & \sum q^{X_2(L)}\prod_{i=1}^{\nu-2}\left[\begin{array}{c}
n_i+\tilde{n}_i-2\delta_{i,\nu-2} \\ n_i-\delta_{i,\nu-2}
\end{array}\right]_q
\label{2.28}
\end{eqnarray}

\noindent
which can be easily verified with an aid of the $q$-binomial recurrences
(\ref{2.5}). Then, using
\begin{equation}
\{\vec{n},\vec{\tilde n}\}_{L,2}=\{\vec{n},\vec{\tilde n}\}_{L-1,3}
\label{2.29}
\end{equation}
and
\begin{equation}
\{\vec{n},\vec{\tilde n}\}_{L-2,2}=\{\vec{n},\vec{\tilde n}\}_{L,2}
-\{\vec{e}_{\nu-2},0\},
\label{2.30}
\end{equation}
one finds that
\begin{equation}
X_2(L)+n_{\nu-2}+n_{\nu-3}=\frac{L}{2}+X_3(L-1)
\label{2.31}
\end{equation}
and
\begin{equation}
X_2(L)=X_2(L-2).
\label{2.32}
\end{equation}

\noindent
Now it is obvious that the first (third) term in (\ref{2.28}) is
$q^{\frac{L}{2}}J_q(L-1,3)(J_q(L-2,2))$. Hence,
equation (\ref{2.28}) becomes
\begin{eqnarray}
J_q(L,2)-q^{\frac{L}{2}}J_q(L-1,3)-J_q(L-2,2)=\;\;\;\;\; \nonumber \\
\sum q^{X_2(L)+n_{\nu-2}}\cdot
\prod_{i=1}^{\nu-2}\left[\begin{array}{c}
n_i+\tilde{n}_i-2\delta_{i,\nu-2}-\delta_{i,\nu-3} \\ n_i-\delta_{i,\nu-3}
\end{array}\right]_q .
\label{2.33}
\end{eqnarray}

\noindent
To proceed further, we shall make the change of summation variables
\begin{eqnarray}
\vec{n} & \rightarrow & \vec{n}+\vec{e}_{\nu-3}-2\vec{e}_{\nu-2} \nonumber \\
\vec{\tilde n} & \rightarrow & \vec{\tilde n}+2\vec{e}_{\nu-2}.
\label{2.34}
\end{eqnarray}

\noindent
This transforms expression (\ref{2.33}) to
\begin{eqnarray}
J_q(L,2)-q^{\frac{L}{2}}\cdot J_q(L-1,3)-J_q(L-2,2)= \nonumber \\
\sum q^{\frac{L-1}{2}+X_1(L-3)}\cdot
\prod_{i=1}^{\nu-2}\left[\begin{array}{c}
n_i+\tilde{n}_i-2\delta_{i,\nu-2} \\ n_i-2\delta_{i,\nu-2}
\end{array}\right]_q .
\label{2.35}
\end{eqnarray}
Making use of
\begin{equation}
\{\vec{n},\vec{\tilde n}\}_{L,2}-\{ 2\vec{e}_{\nu-2},0\}
=\{\vec{n},\vec{\tilde n}\}_{L-3,1}
\label{2.36}
\end{equation}

\noindent
implied by (\ref{2.19}) and (\ref{2.20}), we arrive at a simple form
\begin{equation}
J_q(L,2)-q^{\frac{L}{2}}\cdot J_q(L-1,3)-J_q(L-2,2)=
q^{\frac{L-1}{2}}\cdot J_q(L-3,1).
\label{2.37}
\end{equation}

\noindent
Recalling (\ref{2.27}), we finally have
\begin{equation}
J_q(L,2)-q^{\frac{L}{2}}\cdot J_q(L-1,3)=J_q(L-2,2)+
q^{\frac{L-1}{2}}\cdot\{ J_q(L-1,1)-
q^{\frac{L-1}{2}}\cdot J_q(L-2,2)\} .
\label{2.38}
\end{equation}

\noindent
This completes our proof of (\ref{2.16}) for $r=2$. \\
Let us now highlight the main points of a general proof of
recurrences (\ref{2.15}) and (\ref{2.16}). We start as before, by
expanding $J_q(L,r)(I_q(L,r))$ in a telescopic fashion. We then, recognize
the first term in this expansion as
$q^{\frac{L}{2}}J_q(L-1,r+1)(q^{\frac{L}{2}}I_q(L-1,r-1))$ and the last one
as $J_q(L-2,r)(I_q(L-2,r))$. Subtracting these two terms, we are left with
the sum of $(r-1)((\nu-r-1))$ terms, each one being the sum of
$q$-binomial products. We now relabel the summation variables in a
systematic, but \underline{individual} way (i.e. re-labeling procedure
is different for each term!). At this point it is natural to introduce
interpolating functions $Z_t$ and $\tilde{Z}_t$ which have the
following remarkable properties
\begin{eqnarray}
\left\{\begin{array}{l}
Z_0=J_q(L,r)-\theta(\nu-1>r)\cdot q^{\frac{L}{2}}
\cdot J_q(L-1,r+1)-J_q(L-2,r) \\[2mm]
Z_{r-2}=q^{\frac{L-1}{2}}\{J_q(L-1,r-1)-
q^{\frac{L-1}{2}}\cdot J_q(L-2,r)\} \\[2mm]
Z_t=Z_{t+1}\;;\;\;\;\;\;\;\;\;\;\;t=0,1,2,\ldots,r-3
\end{array}\right.
\label{2.39}
\end{eqnarray}
and
\begin{eqnarray}
\left\{\begin{array}{l}
\tilde{Z}_0=\theta(\nu-1>r)
\{I_q(L,r)-\theta(r>1)\cdot q^{\frac{L}{2}}
\cdot I_q(L-1,r-1)-I_q(L-2,r)\} \\[2mm]
\tilde{Z}_{\nu-2-r}=\theta(\nu-1>r)\cdot
q^{\frac{L-1}{2}}\{I_q(L-1,r+1)-q^{\frac{L-1}{2}}\cdot I_q(L-2,r)\} \\[2mm]
\tilde{Z}_t=\tilde{Z}_{t+1}\;,\;\;\;\;\;\;\;\;\;\;\;\;\;\;\;\;\;\;\;\;
t=0,1,\ldots,\nu-r-3.
\end{array}\right.
\label{2.40}
\end{eqnarray}

\noindent
The equation (\ref{2.39}) and (\ref{2.40}) clearly imply
recurrences (\ref{2.15}) and (\ref{2.16}).
We now refer the motivated reader to the {\em Appendices B} and
{\em C} for a detailed discussion.\\
Reflecting on formulas (\ref{2.15}) and (\ref{2.16}),
one can't help but notice that recurrences for $J_q(L,r)$ and $I_q(L,2)$
are quite similar. To strengthen this similarity we introduce
\begin{equation}
\tilde{I}_q(L,r)\equiv J_q(L+1,r)-
q^{\frac{L+1}{2}}\cdot J_q(L,r+1)\cdot\theta(\nu-1>r)
\label{2.41}
\end{equation}

\noindent
and rewrite equation (\ref{2.16}) as
\begin{equation}
\tilde{I}_q(L,r)=J_q(L-1,r)+\theta(r>1)\cdot
q^{\frac{L}{2}}\tilde{I}_q(L,r-1).
\label{2.42}
\end{equation}

\noindent
Using (\ref{2.41}) and (\ref{2.42}) above, we derive recursion
relations for $\tilde{I}_q(L,r)$
\begin{eqnarray}
\begin{array}{l}
\{\tilde{I}_q(L,r)-\theta(r>1)\cdot q^{\frac{L}{2}}
\cdot\tilde{I}_q(L-1,r-1)\}
-\theta(\nu-1>r)\times \\[2mm]
\times q^{\frac{L-1}{2}}\cdot\{\tilde{I}_q(L-1,r+1)-
q^{\frac{L-1}{2}}\cdot\tilde{I}_q(L-2,r)\}=\tilde{I}_q(L-2,r).
\end{array}
\label{2.43}
\end{eqnarray}

\noindent
Comparing (\ref{2.43}) and (\ref{2.15}) it is plain that $I_q(L,r)$ and
$\tilde{I}_q(L,r)$ satisfy identical recurrences. This fact together
with initial conditions
\begin{eqnarray}
\tilde{I}_q(L=1,r) & = & I_q(L=1,r) \nonumber \\
\tilde{I}_q(L=1,r) & = & I_q(L=0,r)
\label{2.44}
\end{eqnarray}
implies that
\begin{equation}
\tilde{I}_q(L,r)=I_q(L,r) .
\label{2.45}
\end{equation}

\noindent
Hence, we have a $q$-analogue of (\ref{1.54})
\begin{eqnarray}
\left\{ \begin{array}{l}
J_q(L,r)=I_q(L-1,r)+q^{\frac{L}{2}}\cdot J_q(L-1,r+1)\cdot\theta(\nu-1>r) \\
I_q(L,r)=J_q(L-1,r)+q^{\frac{L}{2}}\cdot I_q(L-1,r-1)\cdot\theta(r>1).
\end{array}\right.
\label{2.46}
\end{eqnarray}

\noindent
$J_q(L,r)$ and $I_q(L,r)$ can be fused into a single object
\begin{eqnarray}
F_q(L,r)\equiv \left\{ \begin{array}{lll}
I_q(L,r)\;; & & L+r\equiv odd \\
J_q(L,r)\;; & & L+r\equiv even
\end{array} \right.
\label{2.47}
\end{eqnarray}

\noindent
which satisfies the following recurrences
\begin{eqnarray}
F_q(L,r)=F_q(L-1,r)+q^{\frac{L}{2}} \left\{ \begin{array}{lll}
F_q(L-1,r-1)\cdot\theta(r>1)\;; && L+r\equiv odd \\
F_q(L-1,r+1)\cdot\theta(\nu-1>r)\;; && L+r\equiv even
\end{array} \right.
\label{2.48}
\end{eqnarray}

\noindent
and initial conditions
\begin{eqnarray}
F_q(L=0,r) & = & \delta_{1,r} \nonumber \\
F_q(L=1,r) & = & \delta_{1,r}+q^{\frac{1}{2}}\cdot\delta_{2,r}.
\label{2.49}
\end{eqnarray}

\noindent
Remarkably, recurrences (\ref{2.48}) derived above, are practically
the same as those studied by Andrews, Baxter, and Forrester \cite{17}.
Following along the lines of Schur's polynomial proof of the
Rogers-Ramanujan identities \cite{16}, Andrews, Baxter, and
Forrester introduced another useful representation for the
solution of (\ref{2.48})
\footnote[5]{To make reading easier, we took the liberty to present
results of \cite{17} in the notations of this paper.}
\begin{eqnarray}
B_q(L,r)=\sum_{j=-\infty}^{\infty} \left\{ q^{\varphi_r(j)}
\left[ \begin{array}{c} L \\
((\frac{L+1-r}{2}))-j(\nu+1) \end{array}\right]_q-
q^{\tilde{\varphi}_r(j)}
\left[ \begin{array}{c} L \\
((\frac{L-1-r}{2}))-j(\nu+1) \end{array}\right]_q \right\}
\label{2.50}
\end{eqnarray}
where
\begin{equation}
\varphi_r(j)=\frac{r(r-1)}{4}+j[j\cdot\nu(\nu+1)+r(\nu+1)-\nu]
\label{2.51}
\end{equation}
and
\begin{equation}
\tilde{\varphi}_r(j)= \frac{r(r-1)}{4}+(j\cdot\nu+r)\cdot[j(\nu+1)+1].
\label{2.52}
\end{equation}

\noindent
Making use of $q$-binomial identities (\ref{2.5}, \ref{2.6}), it is a
simple matter to verify that $B_q(L,r)$ satisfy (\ref{2.48}) and (\ref{2.49}).
Since equations (\ref{2.48}) and (\ref{2.49}) specify polynomials
uniquely, we can state the main result of this section
\begin{equation}
F_q(L,r)=B_q(L,r).
\label{2.53}
\end{equation}

\noindent
Result (\ref{2.53}) can be entirely expressed in terms of
$\vec{m}$-variables to obtain a $q$-analogue of (\ref{1.50})
\begin{eqnarray}
\sum_{\vec{V}_r(L)}q^{\frac{\vec{m}^tC\vec{m}}{4}}\cdot
\prod_{i=1}^{\nu-2} \left[ \begin{array}{c}
\frac{1}{2}(K_{\nu-2}\cdot\vec{m}+\vec{u}_r(L))_i+
\frac{L}{2}\delta_{i,\nu-2} \\ m_i
\end{array}\right]_q =\;\;\;\;\;\;\;\;\;\;\;\;\;\;\; \nonumber \\
\sum_{j=-\infty}^{\infty} \left\{
q^{\varphi_r(j)}
\left[ \begin{array}{c} L \\
((\frac{L+1-r}{2}))-j(\nu+1) \end{array}\right]_q -
q^{\tilde{\varphi}_r(j)}
\left[ \begin{array}{c} L \\
((\frac{L-1-r}{2}))-j(\nu+1) \end{array}\right]_q \right\}
\label{2.54}
\end{eqnarray}

\noindent
with Cartan matrix $C$ defined by (\ref{i5}). Once again, the
advantage of representation (\ref{2.54}) is that $\vec{m}$-
variables are practically free, subject only to constraint
(\ref{1.51}).
This fact makes it easy to perform $L\rightarrow\infty$ limit.
We note that the $q$-identities in a form (\ref{2.54}) (modulo minor
notational differences) were conjectured (proven for
$\nu=3,4$) by Melzer in \cite{6}. In my opinion
variables $\vec{n},\vec{\tilde n}$ used
in this paper are more convenient when it comes to the proof.\\
Finally, letting $L$ tend to infinity in (\ref{2.54}), and
using formulas (\ref{2.7}, \ref{2.8}), we complete
the proof of (\ref{i2})
\begin{eqnarray}
\sum^{\infty}_{\stackrel{m_1,m_2,\cdots,m_{\nu-2}=0}
{m_i\equiv V_{i,r}^{\pm}(mod 2)}}
\frac{q^{\frac{\vec{m}^tC\vec{m}}{4}}}{(q,q)_{m_{\nu-2}}}
\prod_{i=1}^{\nu-3} \left[ \begin{array}{c}
\frac{1}{2}(K_{\nu-2}\cdot\vec{m})_i+\frac{1}{2}\delta_{i,b^{\pm}} \\ m_i
\end{array}\right]_q
= \frac{1}{(q,q)_{\infty}}\sum_{j=-\infty}^{\infty}
\{q^{\varphi_r(j)}-q^{\tilde{\varphi}_r(j)}\}
\label{2.55}
\end{eqnarray}
with $b^{\pm}$, defined by (\ref{i3}). \\
To the best of my knowledge, the proof of (\ref{2.55}) is
given here for the first time.\\
Both sums appearing in (\ref{2.55}) have their own technical merits.
In particular, the bosonic sum is best when it comes to the modular
properties; on the other hand, it is the fermionic representation
which enables one to study $q\rightarrow 1^-$ limiting behavior
in terms of Rogers dilogarithms \cite{3} (see also \cite{8}, \cite{10},
\cite{29.5}). By the way, Ramanujan himself
was quite aware of the dilogarithms and knew how to evaluate them
at special points.\\
In the limit $\nu$ tends to infinity, identities (\ref{2.54},\ref{2.55})
imply the new character formula for affine algebra $A_1^{(1)}$ \cite{29.6}.
This formula (which can be interpreted as equivalence of two-particle and
infinite-particle description of XXX-model excitations) plays an important role
in the "Yangian" representation of CFT (\cite{29.7} and references there).\\
To conclude this section, we'd like to point out that one can
generalize polynomial identities (\ref{2.54}) further \cite{6} by adding
to the fermionic system (\ref{1.16}) one more inhomogeneous term. This would
enable us to treat a most general class of boundary conditions
$1 \leq r \leq\nu-1$; $1\leq s \leq\nu$ which correspond to the primary
fields with conformal dimensions $\Delta_{r,s}$ (\ref{i6}).\\
We anticipate that it will be straightforward to extend our
technique to deal with most general boundary conditions.
The details will be given elsewhere \cite{29.8}

\section*{4. Discussion}

\setcounter{chapter}{4}
\setcounter{section}{0}
\setcounter{equation}{0}

In this section we'd like to comment on some "physical" matters
which were left out of the discussion, because of the rather rigid format
of the main body of this paper.\\
To shed even light on the origin of a fermionic (bosonic) sum
terminology, we shall recall two important results from the
theory of partitions \cite{30}. \\
The first one is
\begin{equation}
\frac{1}{(q,q)_{\infty}}=1+\sum_{n=1}^{\infty}p_b(n)\cdot q^n
\label{3.1}
\end{equation}
where $p_b(n)$ is a number of additive partitions on $n$ into
unrestricted number of positive integers which may or may not
be different (order of integers in partition is irrelevant).
Since no exclusion rule is imposed on the parts, it is natural
to refer to $(q,q)_{\infty}^{-1}$ as a bosonic character. \\
The second one deals with a number of additive partitions
$p_f(n,m,N)$ of $n\geq 0$ into $m$ unequal, non-negative
parts which do not exceed $N-1$. This result can be written as
\begin{eqnarray}
\left[\begin{array}{c}
N \\ m \end{array}\right]_q=q^{-\frac{m(m-1)}{2}}
\cdot\sum_{n=0}^{\infty}q^n\cdot p_f(n,m,N).
\label{3.2}
\end{eqnarray}
It is obvious that the exclusion rule for $p_f(n,m,N)$
is of a fermionic nature.\\
We can now give a physical interpretation of phase factors
appearing in (\ref{2.9}, \ref{2.10}). Think of $q$ as a Boltzmann factor
$$ q=e^{-\frac{2\pi v_s}{T\cdot M}}, $$
where $T,M,v_s$ are temperature, length of the system and
speed of sound for the excitations, respectively. Then (\ref{3.2})
implies that a typical term in (\ref{2.9}, \ref{2.10})
\begin{eqnarray}
q^{-\frac{(n_i-\delta_{i,b^{\pm}}-\frac{L}{2}\cdot\delta_{i,\nu-2})
\cdot m_i}{2}}
\left[\begin{array}{c}
n_i+m_i \\ m_i \end{array}\right]_q
\label{3.3}
\end{eqnarray}
is a character (re-scaled partition function) for a massless
system of the $m_i$ fermions with energy $\varepsilon_k$ and
momentum $p_k$, subject to the linear dispersion law
\begin{eqnarray}
\left\{ \begin{array}{l}
\varepsilon_k^{\pm}=v_s\cdot p_k^{\pm}\;;\;\;\;\;\;\;\;\;\;\;
k=0,1,\ldots,n_i+m_i-1 \\
p_k^{\pm}=\frac{2\pi}{M}(k-\frac{n_i+m_i-1}{2}+
\frac{L\cdot\delta_{i,\nu-2}+\delta_{i,b^{\pm}}}{4}) \end{array}\right.
\label{3.4}
\end{eqnarray}

\noindent
and restrictions $$p_{k'}\neq p_k \mbox{  for  } k'\neq k.$$
Equation (\ref{3.4}) shows that the action of $L_0$ (energy) operator takes a
particularly trivial form in fermionic basis. It is natural to pose
a question whether the action of some other operators on fermionic states
can be described in a straightforward fashion. One may hope that the
answer to this question is positive. In fact, it was demonstrated in
the early paper by this author \cite{31} (where $c=1$-model was analyzed)
that the matrix elements of a $U(1)$-current assume a miraculously simple
form in the Bethe (fermionic) basis. That observation made it possible to
establish a map between Virasoro and Bethe states and to obtain
exact results for correlation functions \cite{31}, \cite{31.5}. Provided
one can generalize this approach, it may be possible to study the
finite - volume spectrum of integrable massive perturbations of a rational
CFT analytically, without resorting to the aid of a computer \cite{31.6}. \\
Let us now turn to the most intriguing feature of fermionic sums. They
are not unique. A number of interesting examples was given in \cite{3},
\cite{4}, \cite{5}. In particular, for Ising model one has either $(\nu=3)$
fermionic sum related to a coset construction
$$ \frac{(A_1^{(1)})_1 \otimes(A_1^{(1)})_1}{(A_1^{(1)})_2} $$
or a sum related to construction
$$ \frac{(E_8^{(1)})_1 \otimes(E_8^{(1)})_1}{(E_8^{(1)})_2} $$
with first (second) sum involving one (eight) type(s) of quasi-particles.
Note that both group structures appear naturally in the analysis of
two known integrable perturbations of the Ising model \cite{31.8}.
Another important example, suggesting a possible "infinite degeneracy" of
a fermionic sum description of CFT characters, can be found in \cite{32}.
There it was shown that a simplest bosonic character $(q,q)_{\infty}^{-1}$
can be written in a fermionic language as
$$\frac{1}{(q,q)_{\infty}}=\sum_{m_1,m_2,\ldots,m_{p}\geq 0}
\frac{q^{\sum_{i=1}^p N_i^2}}{(q,q)_{m_p}\cdot
\prod_{i=1}^p(q,q)_{m_i}} $$
where $N_i=\sum_{l=1}^pm_l$. A graph interpretation of a phase factor
$\sum_{i=1}^p N_i^2$ is
$$\sum_{i=1}^p N_i^2 = \vec{m}^t(2-\tilde K)^{-1}\vec{m}$$
with $\tilde K$ being the incidence matrix of a tadpole graph
$\frac{A_{2p}}{Z_2}$ \cite{9}. Most importantly $p$, which can be thought
as a number of different particle types, is an arbitrary positive integer. \\
Now, if one assumes that CFT "knows" about integrable off-critical
extensions, it would be natural to interpret the above mentioned
non-uniqueness of fermionic sums as an indication that there exist many
(perhaps infinitely many) different perturbations of CFT which preserve
integrability \cite{2}-\cite{5}. If it is indeed the case, one should face up
to the possibility that several relevant operators (with finely tuned ratios
of coupling constants) may be added to CFT in such a way that a resulting
theory is still integrable.
It is well established that the information about a mass spectrum and
spins of higher conserved quantities is encoded into the Cartan matrix.
As for the information related to the ratios of relevant operator
coupling constants, translating the phenomenological observations made by
Di Francesco et al \cite{42} into the language of this paper might provide
the necessary insight. \\
Finally I'd like to point out that it is possible to obtain many new
$q$-identities of the Rogers-Ramanujan type through studies of the
higher spin $XXZ$-models in the regime of strong anisotropy. Hopefully
all these challenging questions will be addressed in my future publications.

\section*{Concluding Remark}

I'd like to finish this paper on a personal note.
The author firmly believes that ascertaining the role
of a fermionic-bosonic sum equivalence in the theory of form
factors \cite{43} and in parallel development related
to the $q$-affine algebra \cite{44} would deepen
the synthesis between mathematics and physics,
and would lead to further breakthroughs.

\section*{Acknowledgments}

I'm indebted to B. McCoy and E. Melzer for patient
explanations of their results and for providing me with
many important references.\\
I'd like to thank M. Terhoeven for his collaboration at
the initial stages of this paper and G. Sierra and V. Rittenberg for their
continuous interest. \\
I'm grateful to G. Andrews, R. Flume, I. Frenkel,
G. von Gehlen, W. Nahm, A. Rocha-Caridi, M. R\"{o}sgen,
M. Scheunert, and D. Zagier
for many stimulating discussions, and to the faculty
members of the Physikalisches Institut der Universit\"{a}t Bonn
for the warm hospitality.

\section*{Note added}

After this work was submitted for publication, three papers dealing
with related issues appeared \cite{45}. In the last reference of \cite{45}
it was suggested that identity (\ref{2.54}) proven here implies (via
Bailey-Andrews construction) the validity of similar identities for
characters \\
$\chi_{1,r(k+1)}^{(\nu,k\nu+\nu-1)}(q)$,
$\chi_{1,r(k+1)+k}^{(\nu,k\nu+\nu-1)}(q)$ and
$\chi_{1,kr}^{(\nu,k\nu+1)}(q)$, $\chi_{1,k(r+1)+1}^{(\nu,k\nu+1)}(q)$
for $\nu\geq 4$, $1\leq r\leq\nu-2$ and $k\geq 1$. \\
Finally, I'd like to mention the latest work by A. Kirillov \cite{46} which,
in my opinion, is a "must read" paper for anyone interested in $q$-identities.

\appendix

\section*{Appendix A}

\setcounter{chapter}{1}
\setcounter{equation}{0}

Here, we will establish useful connection between Chebyshev polynomials of
the second kind $U_m$ (\ref{1.11}) and quantities $T_{j,m}$,
defined recursively as
\begin{equation}
T_{j,m}=\frac{T_{j,m-1}}{(1-T_{m,m-1})^{2(j-m)}} \\
\label{a1}
\end{equation}
\begin{equation}
T_{j,0}=x^j
\label{a2}
\end{equation}

\noindent
for $j,m+1=1,2,3,\ldots\;\;$. Substituting
\begin{equation}
T_{1,0}=x=U_1^{-2}
\label{a3}
\end{equation}

\noindent
into (\ref{a1}), we get
\begin{equation}
T_{2,1}=\frac{1}{(\frac{1}{x}-1)^2}=U_2^{-2}.
\label{a4}
\end{equation}

\noindent
Let us now consider the ratio $\frac{T_{j,m}}{T_{j+1,m}}$.
Using (\ref{a1}, \ref{a2}) we find
\begin{eqnarray}
\frac{T_{j,m}}{T_{j+1,m}} & = & (1-T_{m,m-1})^2
\frac{T_{j,m-1}}{T_{j+1,m-1}}=\ldots \nonumber \\
& = & \left(\prod_{l=1}^m(1-T_{l,l-1}) \right) \cdot
\frac{T_{j,0}}{T_{j+1,0}}
=\frac{1}{x}\cdot\prod_{l=1}^m(1-T_{l,l-1}) .
\label{a5}
\end{eqnarray}

\noindent
Note that identity (\ref{a5}) implies that the ratio
$\frac{T_{j,m}}{T_{j+1,m}}$ does not depend on $j$! This observation,
along with two trivial consequences of (\ref{a1})
\begin{eqnarray}
T_{m,m-1} & = & T_{m,m} \nonumber \\
T_{m+2,m} & = & T_{m+2,m+1}\cdot(1-T_{m+1,m})^2 ,
\label{a6}
\end{eqnarray}

\noindent
leads to the following chain
\begin{equation}
\frac{T_{m,m-1}}{T_{m+1,m}}=
\frac{T_{m,m}}{T_{m+1,m}}=
\frac{T_{m+1,m}}{T_{m+2,m}}=
\frac{T_{m+1,m}}{T_{m+2,m+1}(1-T_{m+1,m})^2} .
\label{a7}
\end{equation}

\noindent
Denoting $T_{m,m-1}^{-1}$ as
\begin{equation}
\tilde{U}_m^2=\frac{1}{T_{m,m-1}}
\label{a8}
\end{equation}

\noindent
we obtain from (\ref{a7})
\begin{equation}
\tilde{U}_{m+2}^2 \cdot\tilde{U}_m^2 = (\tilde{U}_{m+1}^2 -1)^2 .
\label{a9}
\end{equation}

\noindent
Extracting the square root and replacing $m$ by $m-1$, expression (\ref{a9})
becomes
\begin{equation}
\tilde{U}_{m+1} \cdot\tilde{U}_{m-1}=(\tilde{U}_m^2 -1) .
\label{a10}
\end{equation}

\noindent
Comparing (\ref{a10}) and (\ref{1.13}) we see that the recurrences for
$\tilde{U}_m$ are identical to those for the Chebyshev polynomials $U_m$.
This fact together with initial conditions (\ref{a3}, \ref{a4}),
implies that
\begin{equation}
\tilde{U}_m=U_m
\label{a11}
\end{equation}
and, therefore,
\begin{equation}
\frac{1}{T_{m,m-1}}=U_m^2 .
\label{a12}
\end{equation}

\noindent
Now using (\ref{a5}) and (\ref{a12}) it is a simple matter to express
all $T_{j,m}$'s in terms of Chebyshev polynomials
\begin{equation}
T_{j,m}=U_{m+1}^{2(m-j)}\cdot U_m^{2(j-m-1)} .
\label{a13}
\end{equation}

\section*{Appendix B}

\setcounter{chapter}{2}
\setcounter{equation}{0}

In this appendix we will prove the following claim for
$r=1,2,\ldots,\nu-1$
\begin{eqnarray}
\{J_q(L,r)-q^{\frac{L}{2}}\cdot
J_q(L-1,r+1)\cdot\theta(\nu-1>r)\}= \;\;\;\;\;\;\;\;\;\; \nonumber \\
J_q(L-2,r)+\theta(r>1)\cdot q^{\frac{L-1}{2}}
\{J_q(L-1,r-1)-q^{\frac{L-1}{2}}\cdot J_q(L-2,r)\}
\label{b1}
\end{eqnarray}

\noindent
where for $L+r\equiv even$ integer
\begin{eqnarray}
J_q(L,r)=q^{X_r(L)}\prod_{i=1}^{\nu-2} \left[\begin{array}{c}
n_i+\tilde{n}_i-V_{i,r} \\ n_i \end{array}\right]_q
\label{b2}
\end{eqnarray}

\noindent
and $V_{i,r}$, $X_r(L)$ and $\{\vec{n},\vec{\tilde n}\}_{L,r}$ were
defined by (\ref{1.20}), (\ref{2.11}), (\ref{1.17}, \ref{1.19})
respectively. For notational simplicity we've suppressed
the summation symbol in (\ref{b2}). Nevertheless, it should
be remembered that the sum over all solutions to
constraint (\ref{1.17}) is always assumed in (\ref{b2}). \\
We start by expanding $J_q(L,r)$ in a telescopic fashion
\[
J_q(L,r)= q^{X_r(L)}\left\{ \theta(\nu-1>r)\cdot
q^{\sum_{\nu-r-1}^{\nu-2}n_i}\cdot\prod_{i=1}^{\nu-2}
\left[\begin{array}{c}
n_i+\tilde{n}_i-V_{i,r+1} \\ n_i \end{array}\right]_q \right. +
\]
\begin{eqnarray}
+\sum_{l=0}^{r-1}q^{\theta(r-1>l)\cdot\sum_{\nu-r+l}^{\nu-2}n_i}
\cdot\prod_{i=1}^{\nu-2}\left.\left[\begin{array}{c}
n_i+\tilde{n}_i-V_{i,r}-\theta(i>\nu-2-r+l) \\
n_i-\delta_{i,\nu-1-r+l}\end{array}\right]_q \right\}.
\label{b3}
\end{eqnarray}

\noindent
To prove formula (\ref{b3}) above, one should simply check
a telescopic properties (\ref{2.17}, \ref{2.18}) using binomial recurrences
(\ref{2.5}). \\
With the aid of easily verifiable identities
\begin{eqnarray}
\{\vec{n},\vec{\tilde n}\}_{L,r} & = &
\{\vec{n},\vec{\tilde n}\}_{L-1,r+1} \nonumber \\
X_r(L)+\sum_{\nu-r-1}^{\nu-2}n_i & = &
\frac{L}{2}+X_{r+1}(L-1)
\label{b4}
\end{eqnarray}

\noindent
implied by (\ref{2.19}), we can immediately identify the first
term in the expansion (\ref{b3}) as \\
$q^{\frac{L}{2}}\cdot J_q(L-1,r+1)\cdot\theta(\nu-1>r)$ and hence,
\begin{eqnarray}
J_q(L,r)-q^{\frac{L}{2}}\cdot J_q(L-1,r+1)\cdot\theta(\nu-1>r)
=q^{X_r(L)}\left\{
\sum_{l=0}^{r-2}q^{\sum_{\nu-r+l}^{\nu-2}n_i}\times
\right. \;\;\;\;\;\;\;\;\;\;\; \nonumber \\[2mm]
\times\prod_{i=1}^{\nu-2} \left.\left[\begin{array}{c}
n_i+\tilde{n}_i-V_{i,r}-\theta(i>\nu-2-r+l) \\
n_i-\delta_{i,\nu-1-r+l}\end{array}\right]_q
+ \prod_{i=1}^{\nu-2}\left[\begin{array}{c}
(n_i-\delta_{i,\nu-2})+\tilde{n}_i-V_{i,r} \\
(n_i-\delta_{i,\nu-2})\end{array}\right]_q \right\} .
\label{b5}
\end{eqnarray}

\noindent
Next, using (\ref{b5}) along with two simple equations below
\begin{eqnarray}
V_{i,r}-\theta(l>1)\sum_{m=0}^{l-2}\delta_{i,\nu-r+m}
& = & V_{i,r-1}+\theta(i>\nu-r-2+l)  \nonumber \\[3mm]
\{\vec{n},\vec{\tilde n}\}_{L-1,r-1} & = &
\{\vec{n},\vec{\tilde n}\}_{L,r}-\{\vec{e}_{\nu-2},0\}
\label{b6}
\end{eqnarray}

\noindent
one derives expansion for
$\{J_q(L-1,r-1)-q^{\frac{L-1}{2}}\cdot J_q(L-2,r)\}$
\begin{eqnarray}
J_q(L-1,r-1) & - & q^{\frac{L-1}{2}}\cdot J_q(L-2,r) =
q^{X_{r-1}(L-1)}\left\{
\sum_{l=1}^{r-2}q^{\sum_{\nu-r+l}^{\nu-2}{n'}_i}
\times \right. \nonumber \\[3mm]
& \times & \prod_{i=1}^{\nu-2} \left[\begin{array}{c}
{n'}_i+\tilde{n}_i-V_{i,r}+\theta(l>1)\cdot
\sum_{m=0}^{l-2}\delta_{i,\nu-r+m} \\
{n'}_i-\delta_{i,\nu-r+l-1}\end{array}\right]_q + \nonumber \\[3mm]
& + & \prod_{i=1}^{\nu-2}\left.\left[\begin{array}{c}
{n'}_i+\tilde{n}_i-V_{i,r}+\sum_{m=0}^{r-3}\delta_{i,\nu-r+m} \\
{n'}_i-\delta_{i,\nu-2} \end{array}\right]_q \right\}
\label{b7}
\end{eqnarray}

\noindent
with ${n'}_i=n_i-\delta_{i,\nu-2}$. This expansion will come in
handy later. \\
Let us now return to expansion (\ref{b5}). Since
\begin{equation}
\{\vec{n},\vec{\tilde n}\}_{L-2,r}=
\{\vec{n},\vec{\tilde n}\}_{L,r}-\{\vec{e}_{\nu-2},0\}
\label{b8}
\end{equation}
and
\begin{equation}
X_r(L-2)=X_r(L)
\label{b9}
\end{equation}

\noindent
we recognize the last term in the r.h.s. of (\ref{b5}) as
$J_q(L-2,r)$. Thus,
\begin{eqnarray}
J_q(L,r)-q^{\frac{L}{2}}\cdot J_q(L-1,r+1)\cdot\theta(\nu-1>r)
-J_q(L-2,r)= \;\;\;\nonumber \\[2mm]
\sum_{l=0}^{r-2}q^{X_r(L)+\sum_{\nu-r+l}^{\nu-2}n_i}\cdot
\prod_{i=1}^{\nu-2}\left[\begin{array}{c}
n_i+\tilde{n}_i-V_{i,r}-\theta(i>\nu-2-r+l) \\
n_i-\delta_{i,\nu-1-r+l}\end{array}\right]_q .
\label{b10}
\end{eqnarray}

\noindent
At this point it is expedient to perform the change
of summation variables in (\ref{b10}). For the $l$-th term, this
change takes the form
\begin{eqnarray}
\vec{n} & \rightarrow & \vec{n}+\vec{e}_{\nu-r+l-1}
-\vec{e}_{\nu-r+l}-\vec{e}_{\nu-2}  \nonumber \\
\vec{\tilde n} & \rightarrow & \vec{\tilde n}+
2\sum_{i=\nu-r+l}^{\nu-2}\vec{e}_i .
\label{b11}
\end{eqnarray}

\noindent
The equation (\ref{b10}) becomes
\begin{eqnarray}
J_q(L,r)-q^{\frac{L}{2}}\cdot J_q(L-1,r+1)\cdot\theta(\nu-1>r)
-J_q(L-2,r)=\;\;\;\;\;\;\;\;\;\;\;\;\; \nonumber \\[3mm]
q^{\frac{L-1}{2}+X_{r-1}(L-1)}\cdot\sum_{l=0}^{r-2}
q^{\theta(l>0)\cdot\sum_{\nu-r}^{\nu-r+l-1}n_i}\cdot
\prod_{i=1}^{\nu-2}\left[\begin{array}{c}
n'_i+\tilde{n}_i-V_{i,r}+\theta(i>\nu-r+l) \\
n'_i-\delta_{i,\nu-r+l}\end{array}\right]_q
\label{b12}
\end{eqnarray}

\noindent
where, once again, $n'_i=n_i-\delta_{i,\nu-2}$. We want to stress
that the change of variables (\ref{b11}) explicitly depends
on $l$, i.e. it is different for each term in the sum appearing
in (\ref{b10}). Next we define for $t=0,1,\ldots,r-2$
\begin{eqnarray}
Z_t &=& q^{\frac{L-1}{2}+X_{r-1}(L-1)} \left\{ \theta(t>0)
\sum_{l'=1}^t q^{\sum_{\nu-r+l'}^{\nu-2}n'_i}
\prod_{i=1}^{\nu-2}\left[\begin{array}{c}
n'_i+\tilde{n}_i-V_{i,r}+\theta(l'>1)
\sum_{m=0}^{l'-2}\delta_{i,\nu-r+m} \\
n'_i-\delta_{i,\nu-r+l'-1}\end{array}\right]_q + \right. \nonumber \\
\label{b13} \\
&+& \sum_{l=t}^{r-2} q^{\theta(l>t)\sum_{\nu-r+t}^{\nu-r+l-1}n'_i}
\prod_{i=1}^{\nu-2}\left.\left[\begin{array}{c}
n'_i+\tilde{n}_i-V_{i,r}+\theta(i>\nu-r+l)+\theta(t>0)
\sum_{m=0}^{t-1}\delta_{i,\nu-r+m} \\
n'_i-\delta_{i,\nu-r+l}\end{array}\right]_q \right\}.  \nonumber
\end{eqnarray}

\noindent
Reflecting upon (\ref{b7}), (\ref{b12}) and (\ref{b13}), one notices
\begin{eqnarray}
Z_0 & = & J_q(L,r)-q^{\frac{L}{2}}\cdot J_q(L-1,r+1)
\cdot\theta(\nu-1>r)-J_q(L-2,r) \nonumber \\[2mm]
Z_{r-2} & = & q^{\frac{L-1}{2}}
\{J_q(L-1,r-1)-q^{\frac{L-1}{2}}\cdot J_q(L-2,r)\} .
\label{b14}
\end{eqnarray}

\noindent
                               \noindent
Identities (\ref{b14}) may be regarded as a motivation for introducing new
objects $Z_t$. To complete the proof of (\ref{b1}) we show for
$t=0,1,\ldots,r-3$
\begin{equation}
Z_t=Z_{t+1} .
\label{b15}
\end{equation}

\noindent
To this end we expand "$l=t$"-term in the r.h.s. of (\ref{b13}) as
\begin{eqnarray}
\prod_{i=1}^{\nu-2}\left[\begin{array}{c}
n'_i+\tilde{n}_i-V_{i,r}+\theta(i>\nu-r+t)+\theta(t>0)
\sum_{m=0}^{t-1}\delta_{i,\nu-r+m} \\
n'_i-\delta_{i,\nu-r+t}\end{array}\right]_q=\;\;\;\;\;\;\;\;\;\; \nonumber \\
\label{b16}\\
\sum_{l=t+1}^{r-1}q^{\theta(l>t+1)\sum_{\nu-r+t+1}^{\nu-r+l-1}n'_i}
\prod_{i=1}^{\nu-2}\left[\begin{array}{c}
n'_i+\tilde{n}_i-V_{i,r}+\theta(i>\nu-r+l)+\theta(t>0)
\sum_{m=0}^{t-1}\delta_{i,\nu-r+m} \\
n'_i-\delta_{i,\nu-r+l}-\delta_{i,\nu-r+t} \end{array}\right]_q . \nonumber
\end{eqnarray}

\noindent
The proof of a telescopic expansion (\ref{b16}) is, by now,
a standard operating procedure, so we leave it as an exercise
for the reader.
Substituting (\ref{b16}) into (\ref{b13}) and, using the
binomial identity (\ref{2.5}), we find
\[
Z_t=q^{\frac{L-1}{2}+X_{r-1}(L-1)} \left\{ \theta(t+1>0)
\sum_{l'=1}^{t+1}q^{\sum_{\nu-r+l'}^{\nu-2}n'_i}
\prod_{i=1}^{\nu-2}\left[\begin{array}{c}
n'_i+\tilde{n}_i-V_{i,r}+\theta(l'>1)
\sum_{m=0}^{l'-2}\delta_{i,\nu-r+m} \\
n'_i-\delta_{i,\nu-r+l'-1}\end{array}\right]_q+ \right.
\]
\[
+\sum_{l=t+1}^{r-2}q^{\theta(l>t+1)
\sum_{\nu-r+t+1}^{\nu-r+l-1}n'_i}
\prod_{\stackrel{i=1}{i\neq \nu-r+t}}^{\nu-2}
\left[\begin{array}{c}
n'_i+\tilde{n}_i-V_{i,r}+\theta(i>\nu-r+l)+\theta(t>0)
\sum_{m=0}^{t-1}\delta_{i,\nu-r+m} \\
n'_i-\delta_{i,\nu-r+l} \end{array}\right]_q \times
\]
\begin{eqnarray}
\times\left( q^{n'_{\nu-r+t}} \left.\left[\begin{array}{c}
n'_{\nu-r+t}+\tilde{n}_{\nu-r+t}-(t+1) \\ n'_{\nu-r+t}
\end{array}\right]_q + \left[\begin{array}{c}
n'_{\nu-r+t}+\tilde{n}_{\nu-r+t}-(t+1) \\ n'_{\nu-r+t}-1
\end{array}\right]_q \right) \right\} =Z_{t+1}\;\;\;
\label{b17}
\end{eqnarray}

\noindent
which proves (\ref{b15}). Now (\ref{b14}) and (\ref{b15})
imply that
\begin{eqnarray}
J_q(L,r)-q^{\frac{L}{2}}\cdot J_q(L-1,r+1)
\cdot\theta(\nu-1>r)-J_q(L-2,r)=Z_0= \nonumber \\
=Z_1=\ldots=Z_{r-2}=q^{\frac{L-1}{2}}\{J_q(L-1,r-1)-
q^{\frac{L-1}{2}}\cdot J_q(L-2,r)\} .
\label{b18}
\end{eqnarray}

\noindent
This completes the proof of our claim (\ref{b1}).

\section*{Appendix C}

\setcounter{chapter}{3}
\setcounter{equation}{0}

Here we will prove the following claim for
$r=1,2,\ldots,\nu-1$
\begin{eqnarray}
\{I_q(L,r)-q^{\frac{L}{2}}\cdot I_q(L-1,r-1)\cdot\theta(r>1)\}=
\;\;\;\;\;\;\;\;\;\;\;\;\;\;\; \nonumber \\
I_q(L-2,r)+\theta(\nu-1>r)\cdot q^{\frac{L-1}{2}}
\{I_q(L-1,r+1)-q^{\frac{L-1}{2}}\cdot I_q(L-2,r)\}
\label{c1}
\end{eqnarray}

\noindent
where for $L+r\equiv odd$ integer
\begin{eqnarray}
I_q(L,r)=q^{Y_r(L)}\prod_{i=1}^{\nu-2} \left[\begin{array}{c}
n_i+\tilde{n}_i+\tilde{V}_{i,r} \\ n_i \end{array}\right]_q
\label{c2}
\end{eqnarray}

\noindent
and $\tilde{V}_{i,r}$, $Y_r(L)$ and $\{\vec{n},\vec{\tilde n}\}_{L,r}$
were defined by (\ref{1.21}), (\ref{2.12}), (\ref{1.17}, \ref{1.19}),
respectively. Once again, the sum over all solutions to
constraint (\ref{1.17}) is assumed in (\ref{c1}). Since the proof
of (\ref{c1}) is very similar to that of (\ref{b1}), we will leave
the details out and will refer the reader to the {\em Appendix B} for
clarification when the need arises. \\
We start by expanding $I_q(L,r)$ in a telescopic fashion (\ref{2.18}).
\begin{eqnarray}
I_q(L,r) & = & q^{Y_r(L)}\left\{ \theta(r>1)\cdot
q^{\sum_{r-1}^{\nu-2}n_i}\cdot\prod_{i=1}^{\nu-2}
\left[\begin{array}{c}
n_i+\tilde{n}_i+\tilde V_{i,r-1} \\ n_i \end{array}\right]_q
\right.+ \nonumber \\[2mm]
& + & \sum_{l=r-1}^{\nu-2}q^{\theta(\nu-2>l)\cdot\sum_{l+1}^{\nu-2}n_i}
\cdot\prod_{i=1}^{\nu-2}\left.\left[\begin{array}{c}
n_i+\tilde{n}_i+\tilde V_{i,r}-\theta(i>l-1) \\
n_i-\delta_{i,l}\end{array}\right]_q \right\}.
\label{c3}
\end{eqnarray}
Observing
\begin{eqnarray}
\left\{ \begin{array}{l}
\{\vec{n},\vec{\tilde n}\}_{L,r}=\{\vec{n},\vec{\tilde n}\}_{L-1,r-1} \\[3mm]
Y_r(L)+\sum_{r-1}^{\nu-2}n_i=\frac{L}{2}+Y_{r-1}(L-1) \end{array}\right.
\label{c4}
\end{eqnarray}
and
\begin{eqnarray}
\left\{ \begin{array}{l}
\{\vec{n},\vec{\tilde n}\}_{L-2,r}=
\{\vec{n},\vec{\tilde n}\}_{L,r}-\{\vec{e}_{\nu-2},0\} \\[3mm]
Y_r(L-2)=Y_r(L), \end{array}\right.
\label{c5}
\end{eqnarray}

\noindent
we identify the first term inside of the figure brackets (\ref{c3})
as $\theta(r>1)\cdot q^{\frac{L}{2}}\cdot I_q(L-1,r-1)$ and
"$l=\nu-2$"-term as $I_q(L-2,r)$. Hence, we have
\begin{eqnarray}
\Delta\equiv I_q(L,r)-\theta(r>1)\cdot q^{\frac{L}{2}}\cdot I_q(L-1,r-1)
-I_q(L-2,r)=\;\;\;\;\;\;\;\;\;\;\;\; \nonumber \\[2mm]
\theta(\nu-1>r)\cdot
\sum_{l=r-1}^{\nu-3}q^{Y_r(L)+\sum_{l+1}^{\nu-2}n_i}\cdot
\prod_{i=1}^{\nu-2}\left[\begin{array}{c}
n_i+\tilde{n}_i+\tilde V_{i,r}-\theta(i>l-1) \\
n_i-\delta_{i,l}\end{array}\right]_q
\label{c6}
\end{eqnarray}
and
\[
I_q(L,r)-\theta(r>1)\cdot q^{\frac{L}{2}}\cdot I_q(L-1,r-1)=
\]
\begin{eqnarray}
\theta(\nu>r)\cdot
\sum_{l=r-1}^{\nu-2}q^{Y_r(L)+\theta(\nu-2>l)\cdot
\sum_{l+1}^{\nu-2}n_i}\cdot
\prod_{i=1}^{\nu-2}\left[\begin{array}{c}
n_i+\tilde{n}_i+\tilde V_{i,r}-\theta(i>l-1) \\
n_i-\delta_{i,l}\end{array}\right]_q .
\label{c7}
\end{eqnarray}

\noindent
Replacing $r$ by $r+1$, $L$ by $L-1$ in (\ref{c7}) and making use of
\begin{eqnarray}
\left\{ \begin{array}{l}
\tilde{V}_{i,r+1}-\theta(i>l-1)=\tilde{V}_{i,r}+\theta(l>r)
\cdot\sum_{m=r}^{l-1}\delta_{i,m} \\[2mm]
\{\vec{n},\vec{\tilde n}\}_{L-1,r+1}=
\{\vec{n},\vec{\tilde n}\}_{L,r}-\{\vec{e}_{\nu-2},0\} \end{array}\right.
\label{c8}
\end{eqnarray}
we find
\[
\theta(\nu-1>r)\{I_q(L-1,r+1)-q^{\frac{L-1}{2}}
\cdot I_q(L-2,r)\} =
\]
\[
=q^{Y_{r+1}(L-1)}\left\{\theta(\nu-2>r)
\sum_{l=r}^{\nu-3}q^{\theta(\nu-2>l)\sum_{l+1}^{\nu-2}{n'}_i}
\cdot\prod_{i=1}^{\nu-2} \left[\begin{array}{c}
{n'}_i+\tilde{n}_i+\tilde{V}_{i,r}+\theta(l>r)
\sum_{m=r}^{l-1}\delta_{i,m} \\
{n'}_i-\delta_{i,l}\end{array}\right]_q + \right.
\]
\begin{eqnarray}
+\theta(\nu-1>r)\cdot\prod_{i=1}^{\nu-2}
\left.\left[\begin{array}{c}
{n'}_i+\tilde{n}_i+\tilde V_{i,r}+\sum_{m=r}^{\nu-3}\delta_{i,m} \\
{n'}_i-\delta_{i,\nu-2} \end{array}\right]_q \right\}
\label{c9}
\end{eqnarray}

\noindent
with ${n'}_i=n_i-\delta_{i,\nu-2}$. Returning to $\Delta$
(\ref{c6}), we perform the change of summation variables,
which for $l$-th term takes the form
\begin{eqnarray}
\vec{n} & \rightarrow & \vec{n}+\vec{e}_l
-\vec{e}_{l+1}-\vec{e}_{\nu-2} \nonumber \\
\vec{\tilde n} & \rightarrow & \vec{\tilde n}+
2\sum_{i=l+1}^{\nu-2}\vec{e}_i ,
\label{c10}
\end{eqnarray}
and obtain
\begin{eqnarray}
\Delta=\theta(\nu-1>r)\cdot
q^{\frac{L-1}{2}+Y_{r+1}(L-1)}\sum_{l=r-1}^{\nu-3}
q^{\theta(l>r-1)\sum_r^l n_i}
\prod_{i=1}^{\nu-2}\left[\begin{array}{c}
n'_i+\tilde{n}_i+\tilde V_{i,r}+\theta(i>l+1) \\
n'_i-\delta_{i,l+1}\end{array}\!\right]_q .
\label{c11}
\end{eqnarray}

\noindent
Next, we introduce for $t=0,1,\ldots,\nu-2-r$ interpolating
function $\tilde Z_t$, defined as
\[
\tilde Z_t = \theta(\nu-1>r)\cdot
q^{\frac{L-1}{2}+Y_{r+1}(L-1)}\times
\]
\[
\times\left\{ \theta(t>0)
\sum_{l'=r}^{r+t-1} q^{\theta(\nu-2>l')
\sum_{l'+1}^{\nu-2}n'_i}
\prod_{i=1}^{\nu-2}\left[\begin{array}{c}
n'_i+\tilde{n}_i+\tilde V_{i,r}+\theta(l'>r)
\sum_{m=r}^{l'-1}\delta_{i,m} \\
n'_i-\delta_{i,l'}\end{array}\right]_q + \right.
\]
\begin{eqnarray}
+\sum_{l=r+t-1}^{\nu-3} q^{\theta(l>r+t-1)\sum_{r+t}^l n'_i}
\prod_{i=1}^{\nu-2}\left.\left[\begin{array}{c}
n'_i+\tilde{n}_i+\tilde{V}_{i,r}+\theta(i>l+1)+\theta(t>0)
\sum_{m=r}^{r+t-1}\delta_{i,m} \\
n'_i-\delta_{i,l+1}\end{array}\right]_q \right\} .
\label{c12}
\end{eqnarray}

\noindent
Inspecting (\ref{c9}), (\ref{c11}) and (\ref{c12}), we notice
\begin{eqnarray}
\tilde Z_0 & = & \Delta  \nonumber \\
\tilde Z_{\nu-2-r} & = & \theta(\nu-1>r)\cdot
q^{\frac{L-1}{2}}\{I_q(L-1,r+1)-q^{\frac{L-1}{2}}
\cdot I_q(L-2,r)\}.
\label{c13}
\end{eqnarray}

\noindent
Following along the lines of {\em Appendix B}, we can prove for
$t=0,1,\ldots,\nu-3-r$
\begin{equation}
\tilde Z_t=\tilde Z_{t+1}.
\label{c14}
\end{equation}
Finally, equation (\ref{c13}) together with (\ref{c14}) implies
\begin{equation}
\Delta=\tilde Z_0=\tilde Z_1=\ldots=
\tilde Z_{\nu-2-r}=\theta(\nu-1>r)\cdot
q^{\frac{L-1}{2}}\{I_q(L-1,r+1)-q^{\frac{L-1}{2}}
\cdot I_q(L-2,r)\}
\label{c15}
\end{equation}

\noindent
which proves the claim (\ref{c1}).

\end{document}